\documentclass[5p,twocolumn]{elsarticle}

\usepackage{amsmath}   
\usepackage{graphicx}  
\usepackage{verbatim}  
\usepackage{color}     
\usepackage{subfigure} 
\usepackage{hyperref}  
\usepackage{graphics}
\usepackage{bm}
\usepackage{ulem}
\usepackage{lineno}
\usepackage{booktabs}
\usepackage{dcolumn}
\usepackage{multirow}
\begin{document}

\title{Evidence for the $N'(1720)3/2^+$ Nucleon Resonance from Combined Studies of CLAS $\pi^+\pi^-p$ Photo-
  and Electroproduction Data}

\author[JLAB]{V.I.~Mokeev\corref{cor1}} \ead{mokeev@jlab.org} 
\author[JLAB]{V.D.~Burkert} 
\author[JLAB]{D.S.~Carman} 
\author[JLAB]{L.~Elouadrhiri}
\author[MSU]{E.~Golovatch} 
\author[SCAROLINA]{R.W.~Gothe}
\author[OHIOU]{K.~Hicks} 
\author[MSU]{B.S.~Ishkhanov} 
\author[MSU]{E.L.~Isupov} 
\author[UCONN]{K.~Joo}   
\author[JLAB,UCONN]{N.~Markov} 
\author[JLAB]{E.~Pasyuk} 
\author[SCAROLINA]{A.~Trivedi}

\cortext[cor1]{Principal corresponding author}

\address[JLAB]{Thomas Jefferson National Accelerator Facility, Newport News, Virginia 23606, USA}
\address[MSU]{Skobeltsyn Institute of Nuclear Physics and Physics Department, Lomonosov Moscow State
  University, 119234 Moscow, Russia}
\address[SCAROLINA]{University of South Carolina, Columbia, South Carolina 29208, USA}
\address[OHIOU]{Ohio University, Athens, Ohio  45701, USA}
\address[UCONN]{University of Connecticut, Storrs, Connecticut 06269, USA}

\begin{abstract} 
  The analysis of the nine 1-fold differential cross sections for the $\gamma_{r,v} p \to \pi^+\pi^-p$ photo- and
  electroproduction reactions obtained with the CLAS detector at Jefferson Laboratory was carried out with the goal
  to establish the contributing resonances in the mass range from 1.6~GeV to 1.8~GeV. In order to describe the photo-
  and electroproduction data with $Q^2$-independent resonance masses and hadronic decay widths in the $Q^2$ range
  below 1.5~GeV$^2$, it was found that an $N'(1720)3/2^+$ state is required in addition to the already well-established
  nucleon resonances. This work demonstrates that the combined studies of $\pi^+\pi^-p$ photo- and electroproduction
  data are vital for the observation of this resonance. The contributions from the $N'(1720)3/2^+$ state and the
  already established $N(1720)3/2^+$ state with a mass of 1.745~GeV are well separated by their different hadronic 
  decays to the $\pi \Delta$ and $\rho p$ final states and the different $Q^2$-evolution of their photo-/electroexcitation
  amplitudes. The $N'(1720)3/2^+$ state is the first recently established baryon resonance for which the results on the 
  $Q^2$-evolution of the photo-/electrocouplings have become available. These results are important for the exploration 
  of the nature of the ``missing'' baryon resonances.       
\end{abstract} 

\begin{keyword} two pion production \sep resonance couplings \sep missing resonances
\PACS{11.55.Fv, 13.40.Gp, 13.60.Le, 14.20.Gk} 
\end{keyword}

\maketitle


\section{Introduction}
\label{intro} 

Studies of the $N^*$ spectrum have been driven for a long time by the search for the so-called ``missing''
baryon states~\cite{Bu16,Cr13,dgirl19,Kl17,Be17}. Different quark models predict many more excited states than
those that have been observed in experiments~\cite{Capst,Giannini:2015zia,hyun}. These predictions rely on the
approximate SU(6) spin-flavor symmetry demonstrated by the pattern of the observed nucleon resonances
\cite{Klempt12}. These model expectations are supported by the studies of the $N^*$ spectrum from the QCD
Lagrangian within Lattice-QCD (LQCD)~\cite{Du12}, consistent with the independent results from continuum QCD
approaches~\cite{Ro11,Cr19s}. In the early few $\mu$s expansion phase of the universe, the transition from a
deconfined mixture of almost massless bare quarks and gauge gluons to a hadron gas of confined quarks and gluons
with dynamically generated masses was mediated by the full spectrum of excited hadrons. This has been
demonstrated in the studies of this phenomenon within LQCD and quark models~\cite{Baz14}. Studies of the $N^*$
spectrum, therefore, address the important open questions on the symmetry of the strong QCD dynamics underlying
nucleon resonance generation and on the emergence of hadronic matter in the universe. 

The data for exclusive meson photoproduction offer a promising avenue in the search for missing resonances
\cite{Bu16,Cr13,dgirl19,Kl17,Be17} through their decays into final states other than the most explored $\pi N$
channel. The missing resonances are expected to have substantial decays to the $\eta N$, $K \Lambda$, $K \Sigma$,
$\pi\pi N$, and $\pi\eta N$ final states accessible in photoproduction, where their photocouplings are expected to
be comparable with those for the observed resonances~\cite{Capst,Giannini:2015zia,Capst1}. Recently, several of the
long-awaited missing resonances were discovered in a global multi-channel analysis of exclusive meson photoproduction
data~\cite{Bu17}, for which the CLAS $KY$ photoproduction data~\cite{Brad06,Brad07,McC09,Dey10} provided a decisive 
impact. Nine new nucleon resonances of three- or four-star status were included in the recent edition of the 
PDG~\cite{Rpp18}. This discovery is consistent with the expectation of SU(6) symmetry in the generation of the $N^*$ 
spectrum. However, this symmetry also predicts many other resonances that have not yet been observed, making a 
continuation of the efforts on the missing resonance search an important avenue in hadron physics.

The CLAS data on exclusive meson electroproduction have extended the capabilities in the search for further missing
resonances~\cite{Mo19,Mo16b,Ri03}. Both the $\pi N$ and $\pi^+\pi^-p$ electroproduction data demonstrate an
increase in the relative resonant contributions with increasing four-momentum transfer $Q^2$~\cite{Bu12,Is16},
making exclusive electroproduction also promising for the exploration of the $N^*$ spectrum.

The opportunities for observation of new nucleon resonances were demonstrated in early analysis of $\pi^+\pi^-p$ 
electroproduction data on three independent 1-fold differential cross sections~\cite{Ri03}. A reasonable description 
of the data in the third resonance region was achieved either by implementing a $N'(1720)3/2^+$ resonance or by 
including only the known resonances but significantly increasing the decay width of the established $N(1720)3/2^+$ to 
the $\pi \Delta$ final state.

In this paper we analyze nine independent $\pi^+\pi^-p$ electroproduction~\cite{Ri03} and photoproduction cross sections 
recently published by the CLAS Collaboration~\cite{Gol19}. A successful description of both of these sets of data with 
$Q^2$-independent resonance masses, and total and partial hadronic decay widths, validates the contributions from resonance states. 
We present evidence for a $N'(1720)3/2^+$ resonance that has been observed together with the known $N(1720)3/2^+$ state 
from combined studies of the CLAS $\pi^+\pi^-p$ photo- and electroproduction data~\cite{Ri03,Gol19} for invariant masses $W$ 
from 1.6 -- 1.8~GeV in the range of photon virtualities $Q^2 < 1.5$~GeV$^2$. The combined studies of the $\pi^+\pi^-p$ photo- 
and electroproduction data provide new clarity that supports the existence of the new $N'(1720)3/2^+$ state. Note that a global 
multi-channel analysis of exclusive meson photoproduction data~\cite{lee13} reports two close resonances with $J^P=3/2^+$ 
spin-parity in the 1.7 -- 1.8~GeV mass range, and that quark models~\cite{Giannini:2015zia,hyun,San15} also predict new 
low-lying baryon states with $J^P=3/2^+$ in this mass interval.

\section{Experimental Data and Analysis Tools}
\label{expanal} 

\begin{figure*}[htb!] 
\begin{center}
\includegraphics[width=0.9\columnwidth,trim=2.2cm 4cm 1.8cm 4.5cm]{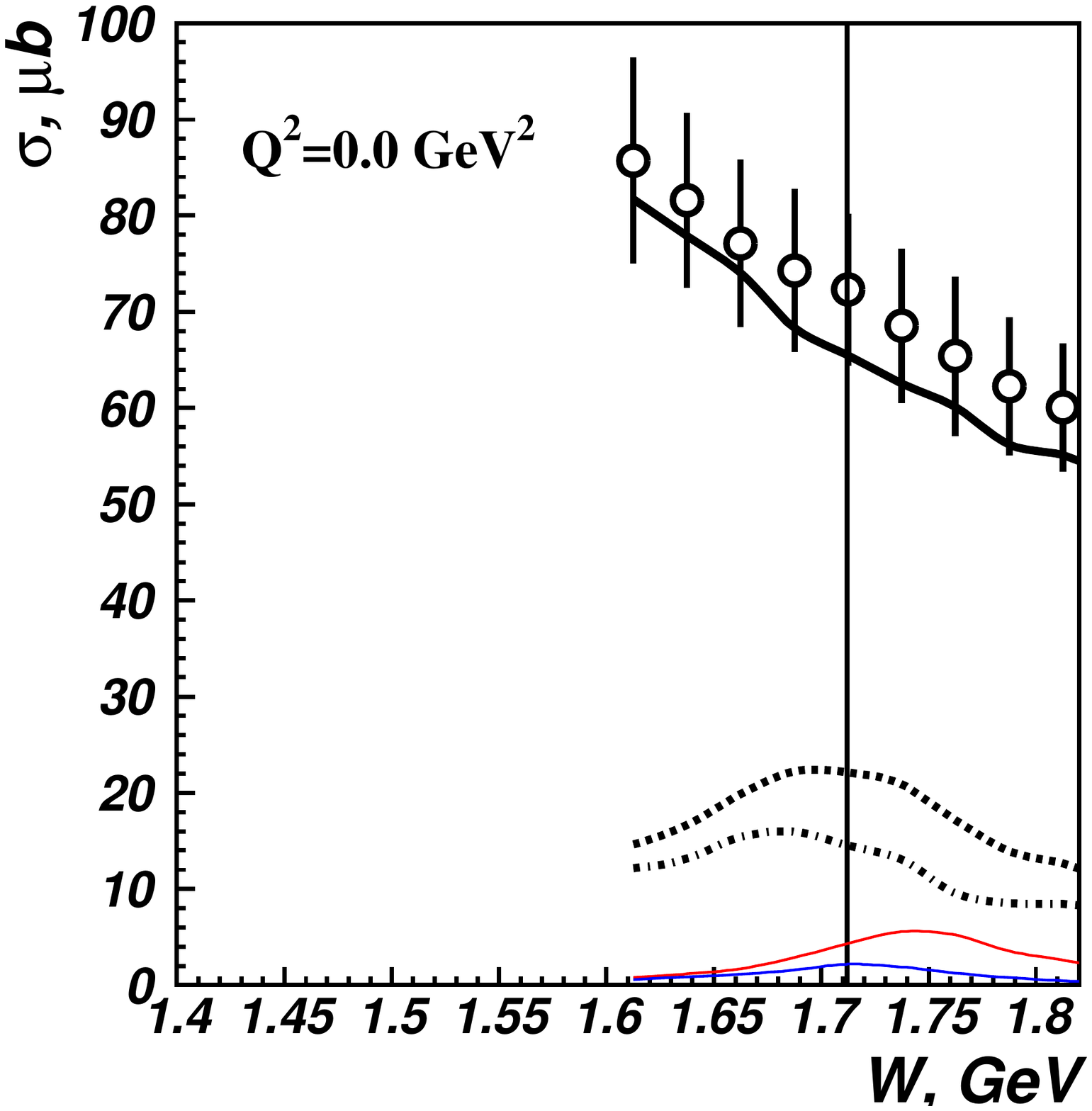}
\includegraphics[width=0.9\columnwidth,trim=2cm 4cm 2cm 4.5cm]{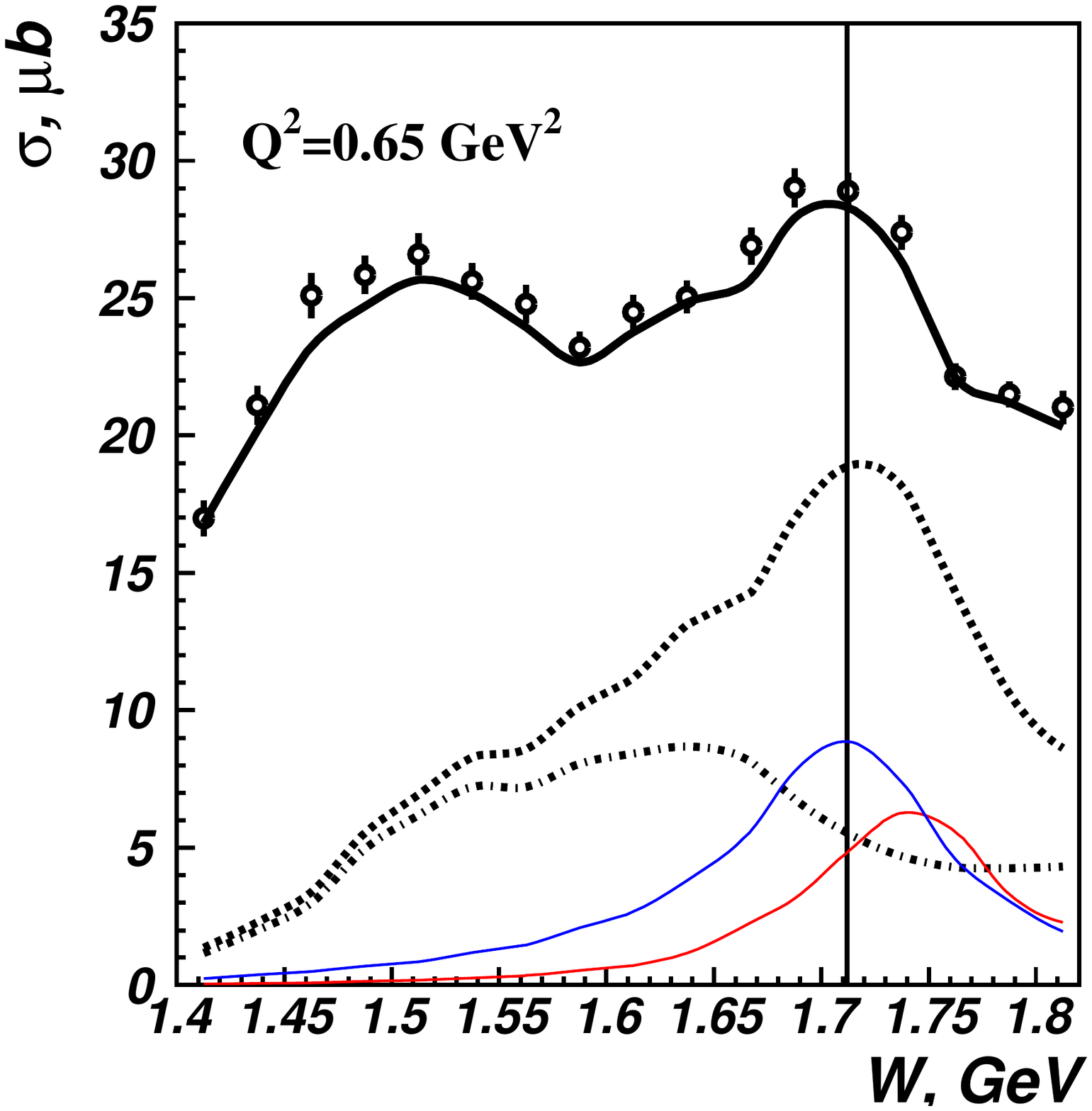}
\includegraphics[width=0.9\columnwidth,trim=2cm 4cm 2cm 4.5cm]{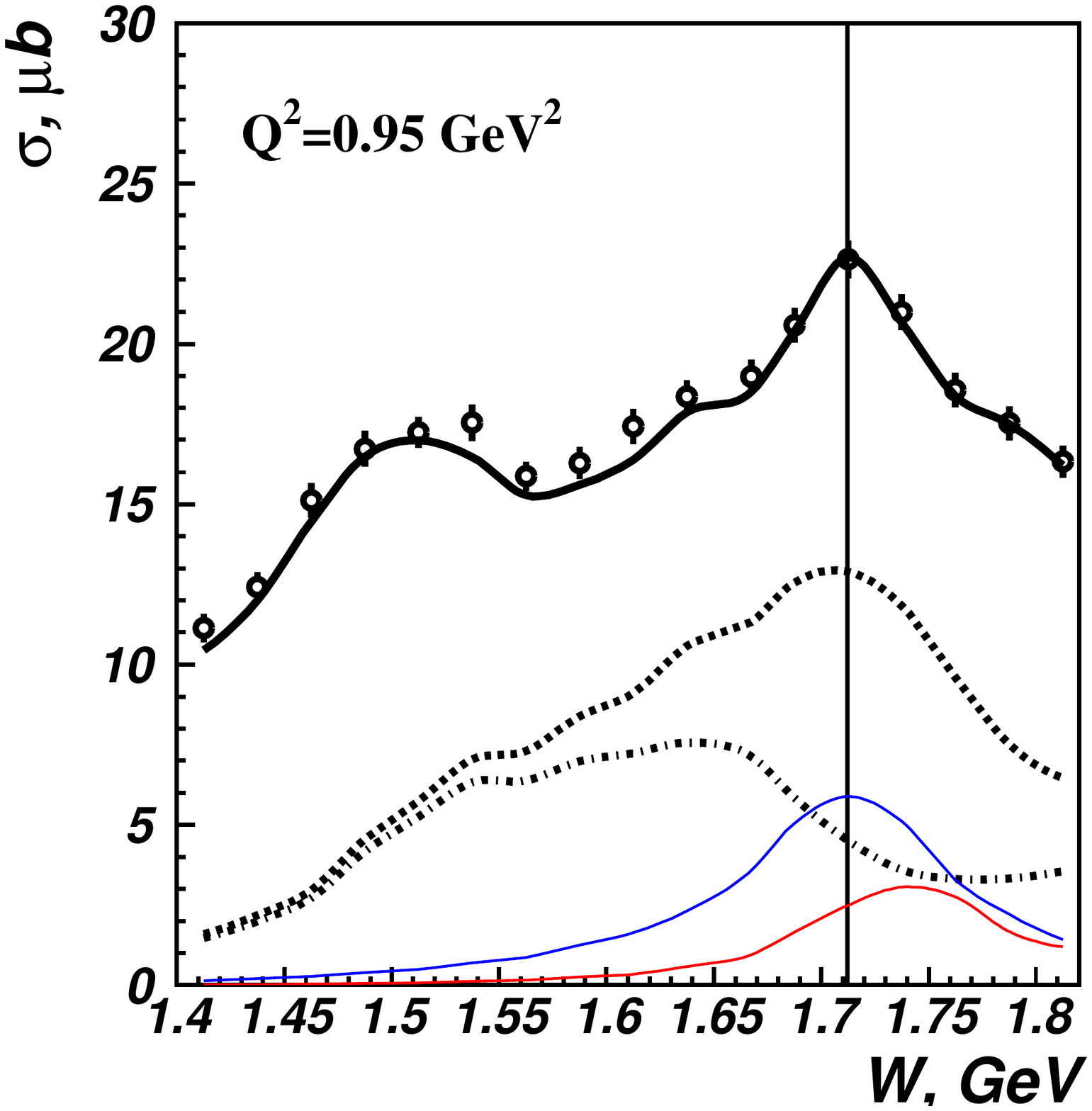}
\includegraphics[width=0.9\columnwidth,trim=2cm 4cm 2cm 4.5cm]{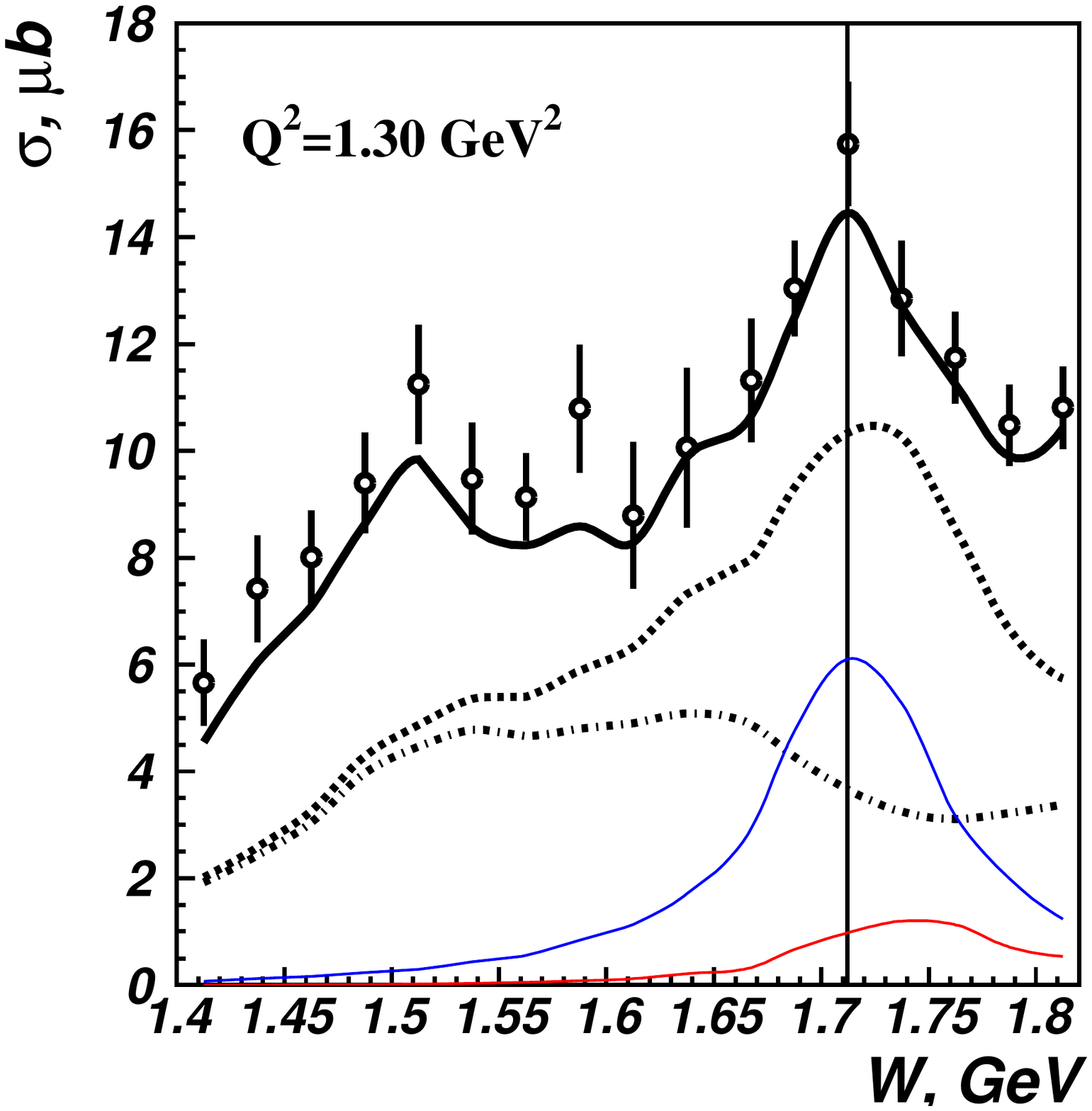}
\vspace{-0.3cm}
\caption{(Color Online) Description of the fully integrated CLAS $\gamma_{r,v} p \to \pi^+\pi^-p'$ photo-/electroproduction
cross sections achieved within the JM model~\cite{Mokeev:2008iw,Mokeev:2012vsa,Mo16a} (shown by the black solid lines). The
error bars include the combined statistical and point-to-point systematic uncertainties for the photoproduction data and only 
the statistical uncertainties for the electroproduction data. The full resonant contributions are shown by the dashed lines
and the dot-dashed lines represent the resonant parts when both the $N(1720)3/2^+$ and $N'(1720)3/2^+$ contributions are taken 
out. The contributions from the $N(1720)3/2^+$ and $N'(1720)3/2^+$ resonances are shown by the thin red and blue lines,
respectively. The vertical lines locate the Breit-Wigner mass of the $N'(1720)3/2^+$ state.}
\label{integ_newres}
\end{center} 
\end{figure*}

In the previous studies of CLAS $\pi^+\pi^-p$ electroproduction data off protons~\cite{Ri03}, two invariant mass
distributions over $M_{\pi^+p}$ and $M_{\pi^+\pi^-}$, and the $\pi^-$ center-of-mass (CM) angular distributions were
analyzed for $W$ from 1.6 -- 1.8~GeV and $Q^2$ from 0.5 -- 1.5~GeV$^2$. A pronounced resonance structure at
$W \approx 1.7$~GeV was observed in all three $Q^2$ bins covered by these data centered at 0.65~GeV$^2$, 0.95~GeV$^2$, 
and 1.3~GeV$^2$ (see Fig.~\ref{integ_newres}). A successful description of these data was only achieved either with a 
much larger branching fraction for the $N(1720)3/2^+$ resonance decays to the $\pi \Delta$ final state in comparison 
with those from experiments with hadron probes or by implementing a new $N'(1720)3/2^+$ baryon state with parameters 
determined from the data fit. Both solutions offered an equally reasonable description of the limited previous CLAS 
$\pi^+\pi^-p$ electroproduction data set. 

In the current analysis we unambiguously established the resonances contributing to $\pi^+\pi^-p$ photo- and
electroproduction in the third resonance region. We have analyzed the data for this channel on the nine 1-fold differential
cross sections and fully integrated cross sections over the final state kinematic variables sorted into nine 25-MeV-wide
bins in $W$ and four $Q^2$-bins at 0~GeV$^2$, 0.65~GeV$^2$, 0.95~GeV$^2$, and 1.30~GeV$^2$. The fully integrated
cross sections and their description achieved within the the JLab-Moscow (JM) meson-baryon model
\cite{Mokeev:2008iw,Mokeev:2012vsa,Mo16a} are shown in Fig.~\ref{integ_newres}.

\begin{figure}[htb!]
\begin{center} 
\includegraphics[width=0.95\columnwidth]{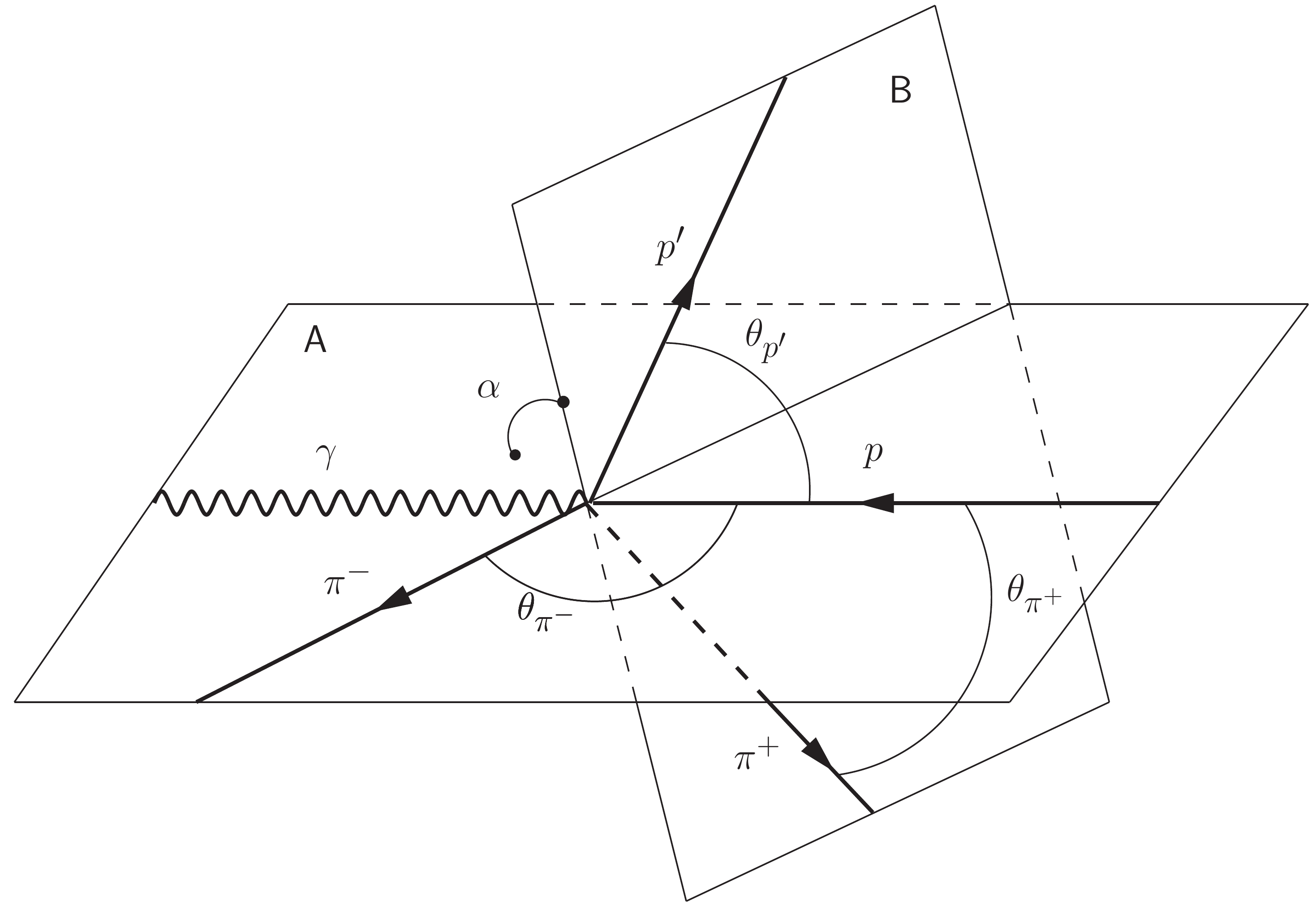} 
\vspace{-0.2cm}
\caption{Angular kinematic variables for the reaction $\gamma p \to \pi^+ \pi^- p'$  in the CM frame. The variable set
with $i$=$\pi^-$, $j$=$\pi^+$, and $k$=$p'$, includes the angular variables for $\theta_{\pi^-}$ (the polar angle of the
$\pi^-$) and $\alpha_{[\pi^-p][\pi^+p']}$, which is the angle between the planes $A$ and $B$, where plane $A$ ($[\pi^-p]$)
is defined by the 3-momenta of the $\pi^-$ and the initial state proton and plane $B$ ($[\pi^+p']$) is defined by the
3-momenta of the $\pi^+$ and the final state proton $p'$. The polar angle $\theta_{p'}$  is relevant for the set with
$i$=$p'$, $j$=$\pi^+$, and $k$=$\pi^-$, while the polar angle $\theta_{\pi^+}$ belongs to the variable set with
$i$=$\pi^+$, $j$=$p'$, and $k$=$\pi^-$.} 
\label{kin}
\end{center} 
\end{figure}

The production of the $\pi^+\pi^-p$ final state hadrons can be fully described by the 5-fold differential cross section
over the invariant masses of the two pairs of the final state hadrons $M_{ij}$, $M_{jk}$ ($i,j,k =\pi^+,\pi^-,p'$) and over
the three angular variables shown in Fig.~\ref{kin}. After integration of the 5-fold differential cross section over the
different sets of four variables, nine 1-fold differential cross sections were determined for:

\begin{enumerate}[a)]
\item Three invariant mass distributions: \newline 
$\frac{d\sigma}{dM_{\pi^+p'}}$, $\frac{d\sigma}{dM_{\pi^+\pi^-}}$, $\frac{d\sigma}{dM_{\pi^-p'}}$;
\item Three angular distributions over $\theta$: \newline 
$\frac{d\sigma}{d(-\cos \theta_{\pi^-})}$, $\frac{d\sigma}{d(-\cos \theta_{\pi^+})}$,
$\frac{d\sigma}{d(-\cos \theta_{p'})}$;
\item Three angular distributions over $\alpha$: \newline 
$\frac{d\sigma}{d\alpha_{[\pi^-p][\pi^+p']}}$, $\frac{d\sigma}{d\alpha_{[\pi^+p][\pi^-p']}}$,
$\frac{d\sigma}{d\alpha_{[p'p][\pi^+\pi^-]}}$.
\end{enumerate}

The chosen nine 1-fold differential cross sections are the most suitable for the exploration of the nucleon resonances
excited in the $\gamma_{r,v}+p$ $s$-channel with subsequent decays into the $\pi\Delta$ and $\rho p$ final states.
The kinematic grid and the number of data points incorporated into the analysis are listed in Table~\ref{exp_data}.
Each of the nine 1-fold differential cross sections, while computed from a common 5-fold differential cross section,
offers complementary information that is essential to gain insight into the resonant contributions.  

The data analysis is carried out within the JM model. This approach incorporates all essential mechanisms seen in the data
that give rise to peaks in the invariant masses and the sharp dependencies in the angular distributions. Less pronounced
mechanisms were established from the correlations between their contributions into the different 1-fold differential
cross sections. The full $\gamma_{r,v} p \to \pi^+\pi^- p'$ amplitudes are described in the JM model as a superposition
of the $\pi^-\Delta^{++}$, $\pi^+\Delta^0$, $\rho p$, $\pi^+ D_{13}^0(1520)$, and $\pi^+ F_{15}^0(1685)$ sub-channels
with subsequent decays of the unstable hadrons to the final state, and direct 2$\pi$ production mechanisms, where the
reaction does not go through the intermediate process of forming unstable hadrons. The JM model incorporates
contributions from all well-established $N^*$ states with observed decays into the $\pi\Delta$ and $\rho p$ final states
listed in Refs.~\cite{Gol19,Mo16a}. For the resonant amplitudes, a unitarized Breit-Wigner ansatz is employed, which makes the
resonant amplitudes consistent with restrictions imposed by a general unitarity condition~\cite{Mokeev:2012vsa}.

The good description of the nine 1-fold differential $\pi^+\pi^-p$ photo- and electroproduction cross sections, achieved 
within the JM model for $W < 2.0$~GeV and in the $Q^2$-range up to 5.0~GeV$^2$, allows isolation of the resonant 
contributions~\cite{Mo19,Gol19,Mo16a} necessary for the extraction of the resonance parameters. The $N^*$ 
photo-/electroexcitation amplitudes ($\gamma_{r,v}pN^*$ photo-/electrocouplings) were determined from analyses of the 
$\pi^+\pi^-p$ photo-/electroproduction data for the resonances in the mass range up to 2.0~GeV from the photoproduction 
data and up to 1.8~GeV from the electroproduction data. Consistent results on the electrocouplings of the $N(1440)1/2^+$ 
and $N(1520)3/2^-$ resonances in the $Q^2$-range from 0.2~GeV$^2$ to 5.0~GeV$^2$ from independent analyses of the dominant 
$\pi N$ and $\pi^+\pi^-p$  electroproduction channels validates the extracted $N^*$ parameters from the JM model. The 
photocouplings of most resonances in the mass range from 1.6--2.0~GeV, their hadronic decays to the $\pi \Delta$ and $\rho p$ 
final states, as well as the electrocouplings of several resonances determined within the JM model are included in the 
PDG~\cite{Rpp18}.

\begin{table*}
\begin{center}
  \caption{The kinematic grid and the number of data points for the 1-fold $\pi^+\pi^-p$ differential
    photo-/electroproduction cross sections used in the analysis within the JM model.}
\label{exp_data} 
\vspace{2mm}
\begin{tabular}{|c|c|c|} \hline
\multirow{4}{*}{$Q^2$, GeV$^2$}    & Number of bins over                                     & Total number \\
                    & $W$, $M_{\pi^-\pi^+}$,  $M_{\pi^+ p'}$,  $M_{\pi^- p'}$  & of data points\\
                   &    $\theta_{\pi^-}$,  $\theta_{\pi^+}$,  $\theta_{p'}$ & for $\pi^+\pi^-p$   \\ 
  &    $\alpha_{[\pi^-p][\pi^+p']}$, $\alpha_{[\pi^+p][\pi^-p']}$,  $\alpha_{[pp'][\pi^-\pi^+]}$ &  \\ \hline
                 & 9,  16, 16, 16,                          &    \\
     0.0              &     14, 14, 14,  &  1188   \\
                   &     14, 14, 14   &         \\ \hline
                   & 9,  10, 10, 10, &     \\
   0.65\,, 0.95\,, 1.30            &     10, 10, 10, &  2025     \\
                &      5,  5,  5  &   \\ \hline
\end{tabular}

\end{center}
\end{table*}

\begin{figure*}[htb!] 
\begin{center}
\includegraphics[width=0.8\columnwidth,trim=2.2cm 4cm 1.8cm 4.5cm]{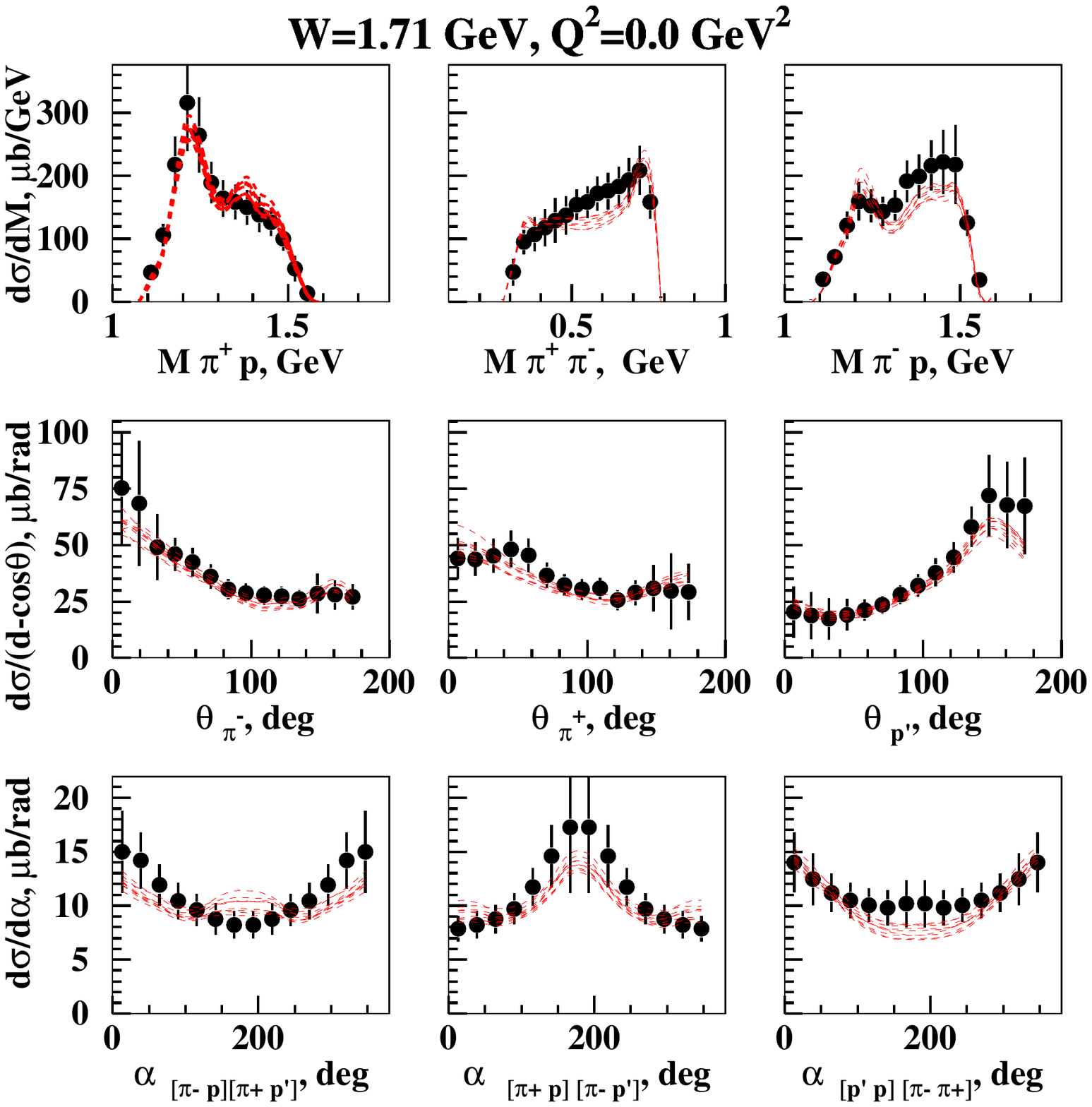}
\includegraphics[width=0.8\columnwidth,trim=2.2cm 4cm 1.8cm 4.5cm]{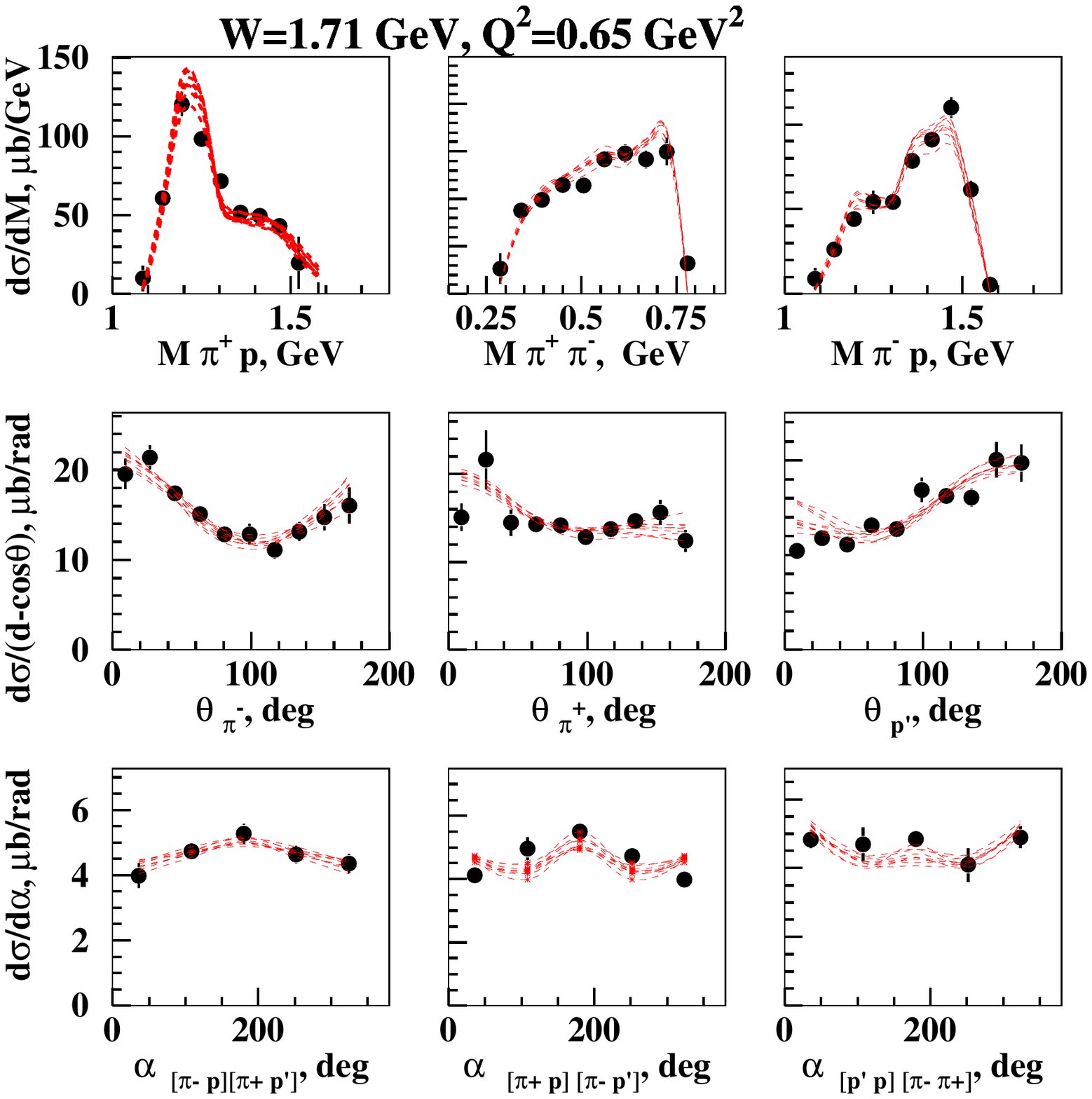}
\includegraphics[width=0.8\columnwidth,trim=2.2cm 4cm 1.8cm 4.5cm]{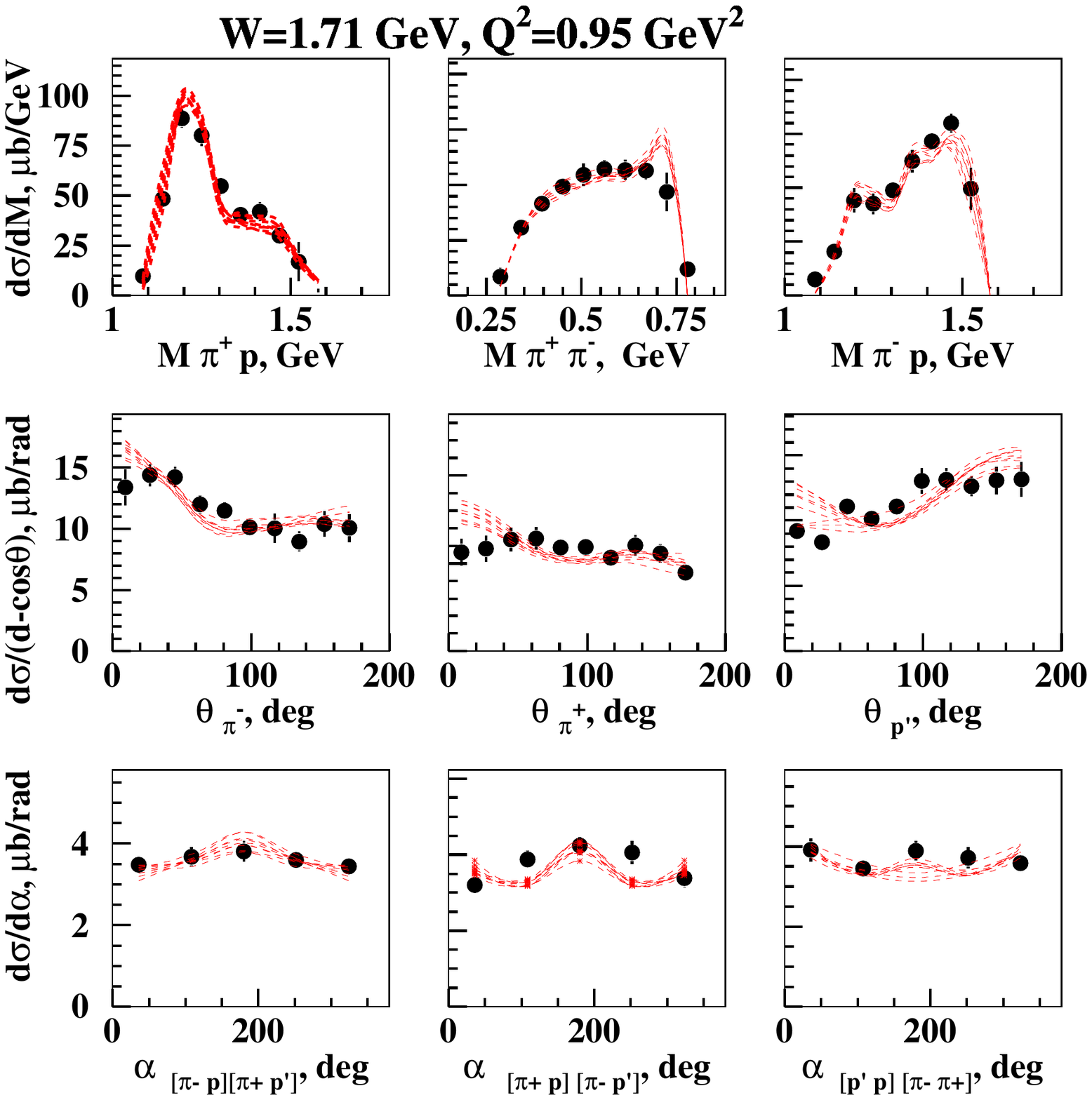}
\includegraphics[width=0.8\columnwidth,trim=2.2cm 4cm 1.8cm 4.5cm]{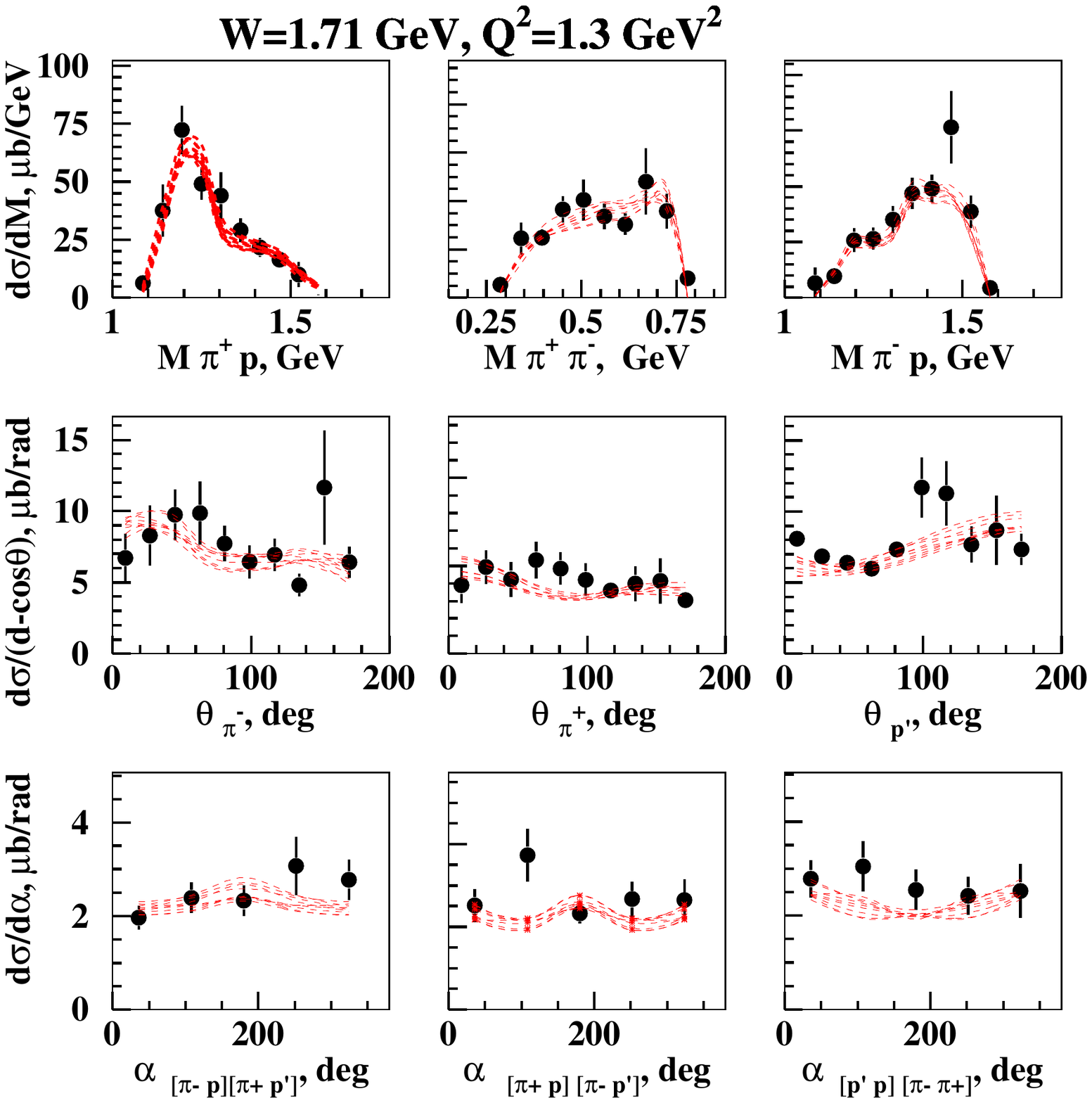}
\vspace{-0.4cm}
\caption{(Color Online) Representative examples for the description of the  $\gamma_{r,v} p \to \pi^+\pi^-p'$ nine
  1-fold differential cross sections achieved within the JM model~\cite{Gol19,Mokeev:2012vsa,Mo16a} for photo- and 
  electroproduction (red curves) in comparison with the data~\cite{Ri03,Gol19}. The error bars include the combined
  statistical and point-to-point systematic uncertainties for the photoproduction data and only the statistical uncertainties
  for the electroproduction data. The group of curves on each plot correspond to the computed cross sections selected in the 
  data fit with $\chi^2/d.p. <  \chi^2/d.p.^{max}$ (see Section~\ref{newres} for details). Fits with $Q^2$-independent masses, 
  and total and partial decay widths into the $\pi\Delta$ and $\rho p$ final states, for all contributing resonances become 
  possible only after the implementation of the new $N'(1720)3/2^+$ state.}
\label{1diff_newres}
\end{center} 
\end{figure*}

\begin{figure*}[htb]
\begin{center}
\includegraphics[width=0.66\columnwidth]{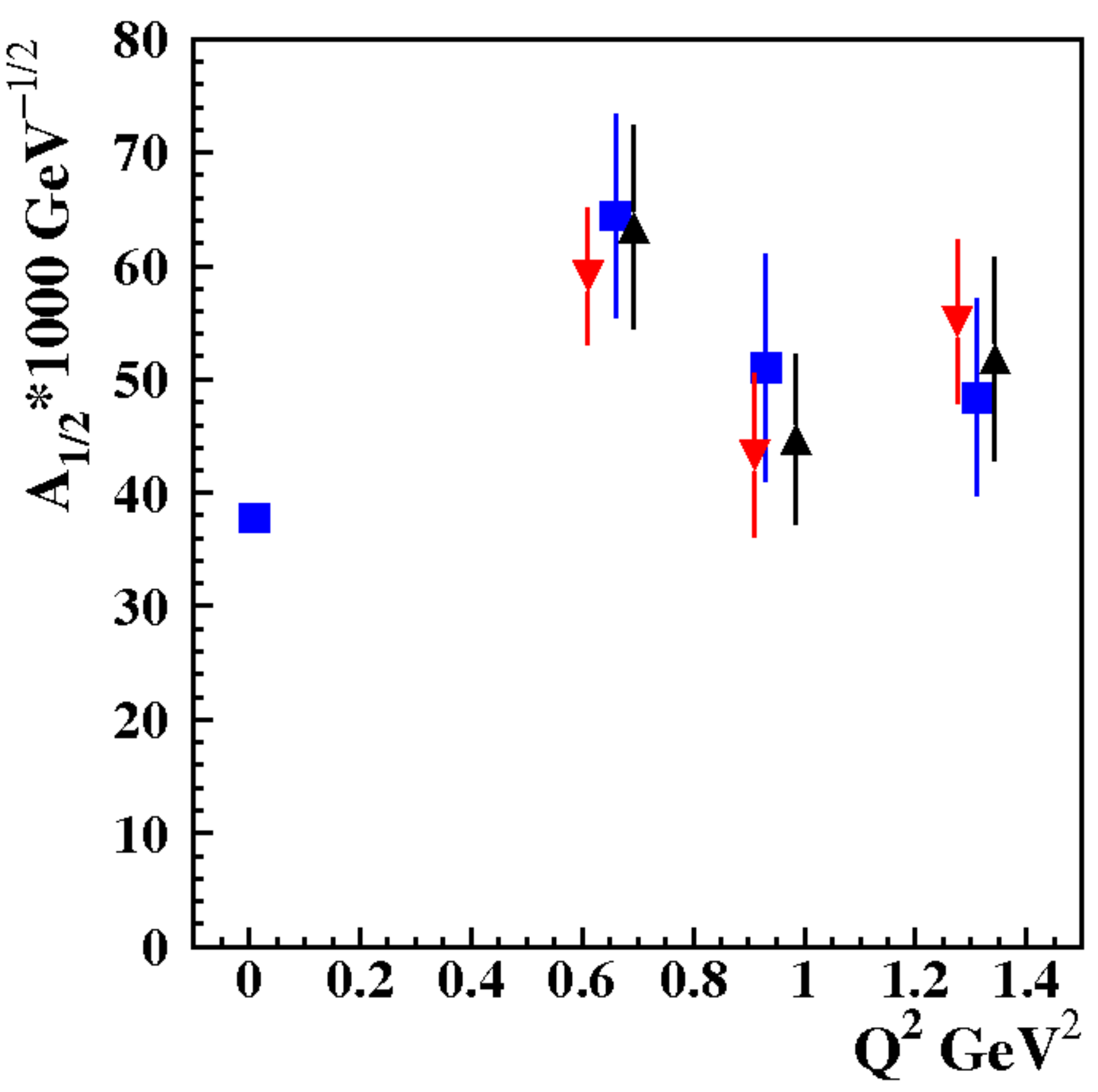}
\includegraphics[width=0.66\columnwidth]{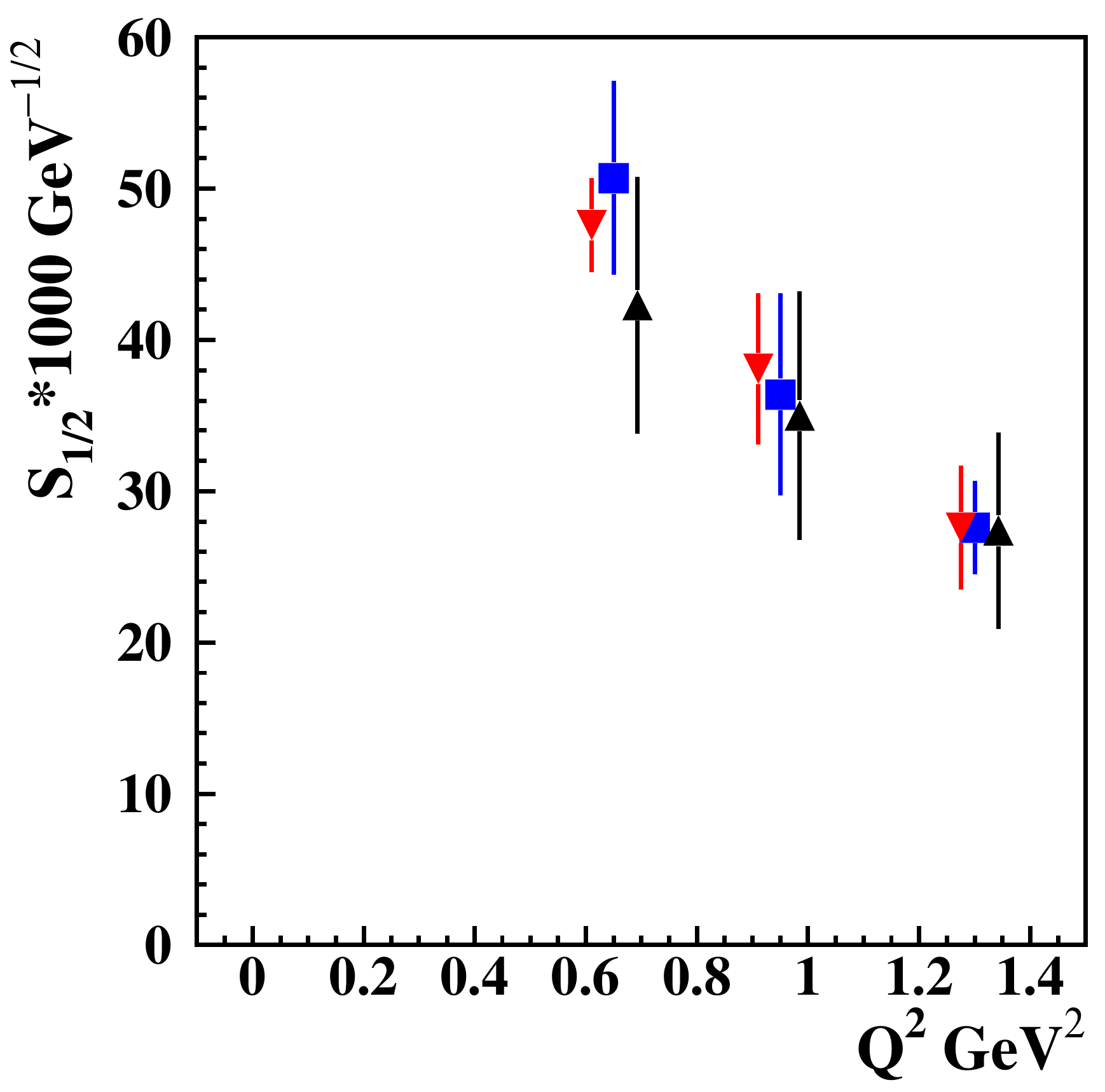}
\includegraphics[width=0.66\columnwidth]{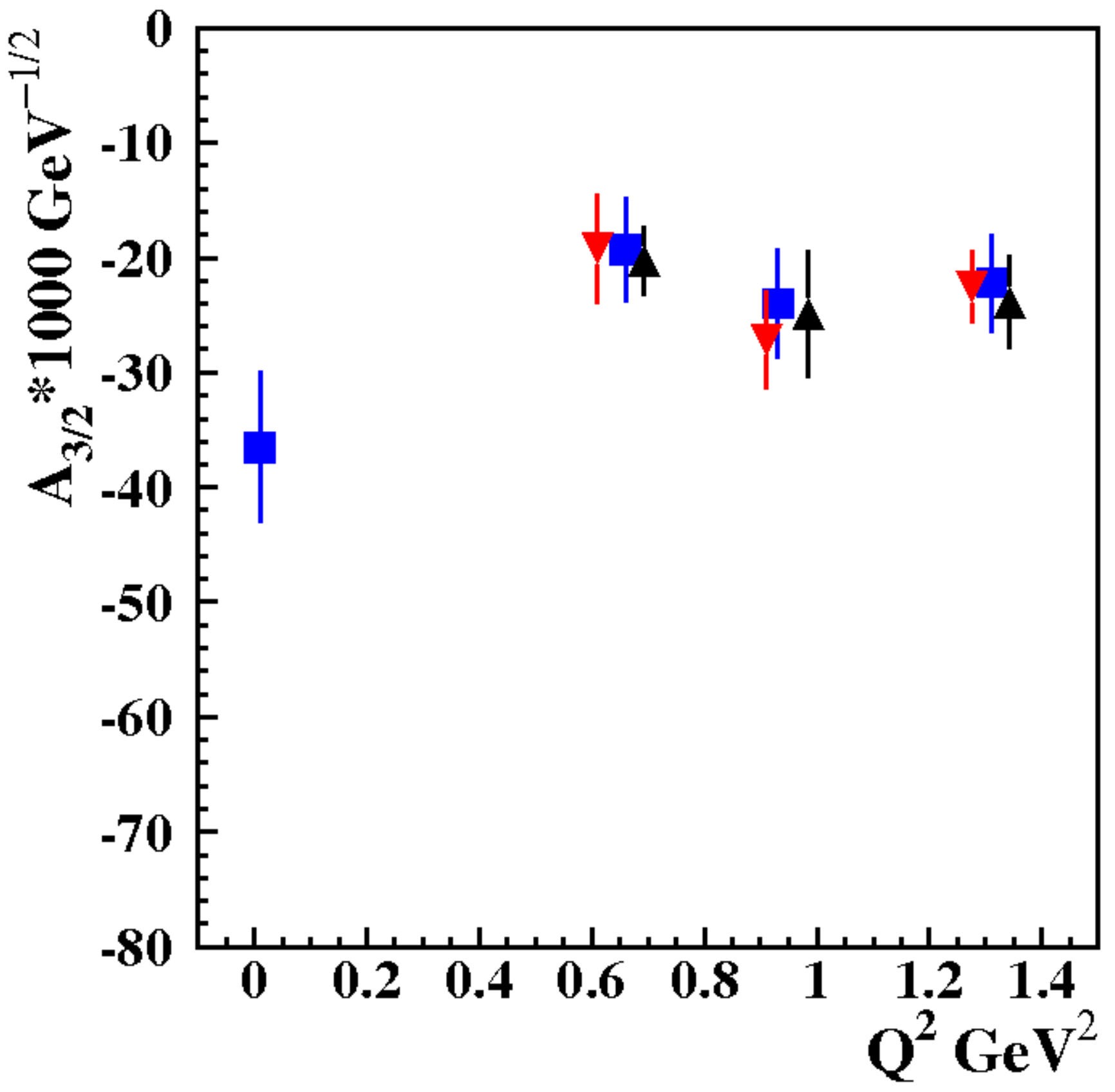}
\vspace{-0.3cm}
\caption{(Color Online) Photo-/electrocouplings of the new $N'(1720)3/2^+$ state determined from independent
  analyses of three overlapping $W$-intervals: a) from 1.61 -- 1.71~GeV (red triangles), b) from 1.66 -- 1.76~GeV 
  (blue squares), and c) from 1.71 -- 1.81~GeV (black triangles). The blue squares at the photon point show the 
  results of the CLAS charged double-pion photoproduction data analysis~\cite{Gol19}.}
\label{new_elecph}
\end{center}
\end{figure*}

\section{Evidence for the New $N'(1720)3/2^+$ Resonance in the $\pi^+\pi^-p$ Data} 
\label{newres}

\begin{table*}[htb!]
\begin{center}
\caption{$N$(1720)3/2$^+$ hadronic decay widths and branching fractions into $\pi \Delta$ and $\rho p$
  determined from independent fits to the data on charged double-pion photo-~\cite{Gol19} and electroproduction
  \cite{Ri03} off protons accounting only for contributions from previously known resonances.}
\label{hadrp13phel}
\vspace{2mm}
\begin{tabular}{|c|c|c|c|} \hline
 \multirow{2}{*}{$N$(1720)3/2$^+$} & $N^*$ total width       & Branching fraction            & Branching fraction         \\
               & MeV                     &  for decays to $\pi\Delta$    &  for decays to $\rho N$    \\ \hline
 electroproduction    &  126.0 $\pm$ 14.0       &    64\% - 100\%               & $<$5\%                    \\
 photoproduction      &  160.0 $\pm$ 65.0       &   14\% - 60\%                 &  19\% - 69\%               \\
\hline
\end{tabular}
\end{center}
\end{table*}

The previous studies of $\pi^+\pi^-p$ photo-/electroproduction with CLAS demonstrated a substantial decrease of the 
relative non-resonant contributions with increasing $Q^2$~\cite{Gol19,Mo16a}. In the current studies, the resonant 
structure clearly seen in $\pi^+\pi^-p$ electroproduction at $W \approx 1.7$~GeV is not visible in the photoproduction 
data because of the maximal non-resonant contributions at $Q^2=0$~GeV$^2$ (see Fig.~\ref{integ_newres}). On the other 
hand, the six angular distributions are sensitive to the resonant contributions both in the photo- and electroproduction 
data. The resonance decays into the $\pi\Delta$ and $\rho p$ final states have a substantial impact on the three invariant 
mass distributions shown in Fig.~\ref{1diff_newres}. The $Q^2$-dependence of the $\pi^+\pi^-p$ photo-/electroproduction 
amplitudes are defined by the $Q^2$-evolution of the nucleon resonance photo-/electrocouplings and the real/virtual 
photon+hadron vertices in the non-resonant mechanisms. However, the resonance masses, as well as their total and partial 
hadronic decay widths, should remain the same in all $Q^2$-bins, as was observed in the analyses of all exclusive meson 
electroproduction data from CLAS~\cite{Mo19,Bu12,Mokeev:2012vsa,Mo16a}. This makes a combined analysis of the $\pi^+\pi^-p$ 
photo-/electroproduction data of particular importance for establishing the resonances contributing to the $\pi^+\pi^-p$ 
channel.

The analyses of the CLAS $\pi^+\pi^-p$ photo- \cite{Gol19} and electroproduction~\cite{Ri03} data were carried out
within the recent version of the JM model in the $W$-range from 1.6 -- 1.8~GeV and for $Q^2$ from 0.0 -- 1.5~GeV$^2$ 
with the goal to establish the resonances contributing in the third resonance region. For the $N(1440)1/2^+$, 
$N(1520)3/2^-$, and $\Delta(1620)1/2^-$ resonances, the initial values of their $\pi\Delta$ and $\rho p$ decay widths 
were taken from analyses of the CLAS $\pi^+\pi^-p$ electroproduction data~\cite{Mokeev:2012vsa,Mo16a}, while for the 
other resonances we used the total hadronic decay widths from the PDG~\cite{Rpp18} and the branching fractions for their 
decays into the $\pi\Delta$ and $\rho p$ final states from Ref.~\cite{Man92}. The initial values of the resonance 
photo-/electrocouplings were taken from the parameterization in Ref.~\cite{Ast} of the available CLAS/world data results 
detailed in Ref.~\cite{webIsupov}. The starting values of the photo-/electrocouplings and the decay widths into the 
$\pi\Delta$ and $\rho p$ final states for the new $N'(1720)3/2^+$ resonance were taken from Refs.~\cite{Ri03,Gol19}.

In the data fit we simultaneously varied the resonance photo-/electrocouplings, the $\pi \Delta$ and $\rho p$ decay
widths, and the parameters of the non-resonant amplitudes described in Refs.~\cite{Gol19,Mokeev:2012vsa,Mo16a}.
$Q^2$-independent hadronic decay widths for all resonances were imposed in the fit. The parameters of the resonant
and non-resonant mechanisms of the JM model were sampled around their initial values, employing unrestricted normal
distributions with a width ($\sigma$) of 30\% of their initial values. In this way, the JM model provided a description
of the observables within their uncertainties for most of the data points. For each trial set of the fit parameters, we
computed the nine 1-fold differential $\pi^+\pi^-p$ cross sections and estimated the $\chi^2/dp$ ($dp$ $\equiv$
data point) values in point-by-point comparisons. We selected the computed cross sections from the data fit within the
range $\chi^2/dp <  \chi^2/dp^{max}$, where $\chi^2/dp^{max}$ was determined so that the computed cross sections
were within the data uncertainties for most data points. The mean values and RMS widths for the resonance parameters
obtained from the fit were used as estimates of their central values and their corresponding uncertainties.

Analyses of the $\pi^+\pi^-p$ photo- and electroproduction data were carried out independently. The resonance
photo-/electrocouplings and the total, $\pi \Delta$, and $\rho p$ decay widths were inferred from the fits over $W$
from 1.6 -- 2.0~GeV for the photoproduction data, and over $W$ from 1.6 -- 1.8~GeV and $Q^2$ from 0.5 -- 1.5~GeV$^2$ 
for the electroproduction data. In the evaluation of $\chi^2/dp$ for the electroproduction data, only the statistical 
uncertainties were taken into account, while for the photoproduction data, the combined contribution from the statistical 
and point-to-point systematic uncertainties was employed since the systematic uncertainties dominate the accuracy of the 
photoproduction data. Parity conservation imposes the requirement of equal values of the $\frac{d\sigma}{d\alpha_i}$ 
($i = \pi^+, \pi^-, p'$) cross sections at the angles $\alpha_i$ and $2\pi - \alpha_i$. This symmetry requirement was 
accounted for in the computation of these cross sections within the JM model. The departure of the data points from this 
requirement, seen only in the $Q^2$=1.3 GeV$^2$ bin, was taken into account in the evaluation of $\chi^2/dp$.

A successful description of the angular $\theta_i$ ($i = \pi^+, \pi^-, p'$) distributions at $W \approx 1.7$~GeV requires 
substantial resonant contributions with $J^P=3/2^+$ spin-parity. This is consistent with the results from previous studies
\cite{Ri03}. The essential role of the $J^{P}=3/2^+$ spin-parity excited nucleon states in the generation of the resonant 
contributions in the third resonance region can be seen in Fig.~\ref{integ_newres}, where the resonant contributions in the 
fully integrated $\pi^+\pi^-p$ photo-/electroproduction cross sections are presented with all relevant resonances included 
and when the contributions from the $N(1720)3/2^+$ and $N'(1720)3/2^+$ resonances are taken out.

We performed two different fits using: 

\begin{table*}[htb]
\begin{center}
  \caption{Hadronic decays into the $\pi \Delta$ and $\rho p$ final states of the resonances in the third resonance region
    with major decays to the $\pi^+\pi^- p$ final state determined from the fits to the data on charged double-pion photo-
    \cite{Gol19} and electroproduction~\cite{Ri03} implementing a new $N '(1720)3/2^+$ baryon state.}
\label{hadr_phot_miss}
\vspace{2mm}
\begin{tabular}{|c|c|c|c|} \hline
Resonance            & $N^*$ total width         & Branching fraction            & Branching fraction         \\
states                & MeV                       & for decays to $\pi\Delta$     &  for decays to $\rho p$    \\ \hline
 $\Delta(1700)3/2^-$  &                           &                               &                           \\
 electroproduction    &  288.0 $\pm$ 14.0         &    77 - 95\%                    &   3 - 5\%                      \\
 photoproduction      &  298.0 $\pm$ 12.0         &    78 - 93\%                    &   3 - 6\%                    \\ \hline
 N(1720)3/2$^+$       &                           &                               &                           \\
 electroproduction    &  116.0 $\pm$ 7.0          &    39 - 55\%                    &   23 - 49\%                     \\
 photoproduction      &  112.0 $\pm$ 8.0          &    38 - 53\%                    &   31 - 46\%                    \\ \hline
N$^{\, '}$(1720)3/2$^+$ &                          &                               &                           \\
 electroproduction    &  119.0 $\pm$ 6.0         &     47 - 64\%                  &    3 - 10\%                     \\
 photoproduction      &  120.0 $\pm$ 6.0         &     46 - 62\%                   &    4 - 13\%                    \\ \hline
\end{tabular}
\end{center}
\end{table*}

\begin{itemize}
\item the contributions from only well-established resonances listed in Refs.~\cite{Gol19,Mo16a}, including the $N(1700)3/2^-$ and $N(1720)3/2^+$ states 
(fit A);
\item the fit A assumptions also adding a new $N'(1720)3/2^+$ resonance with mass, total, $\pi\Delta$, and $\rho p$
decay widths, and photo-/electrocouplings fit to the data (fit B). 
\end{itemize}

Both fits result in good descriptions of the CLAS data with a comparable quality for photo- and electroproduction.
Representative examples of the description of the nine 1-fold differential cross sections with fit B are shown in
Fig.~\ref{1diff_newres}, where the cross sections from the fits within the range $\chi^2/dp <  \chi^2/dp^{max}$
are shown by the family of curves overlaid on each plot.

\begin{figure*}[htb!]
\begin{center}
\includegraphics[width=0.63\columnwidth]{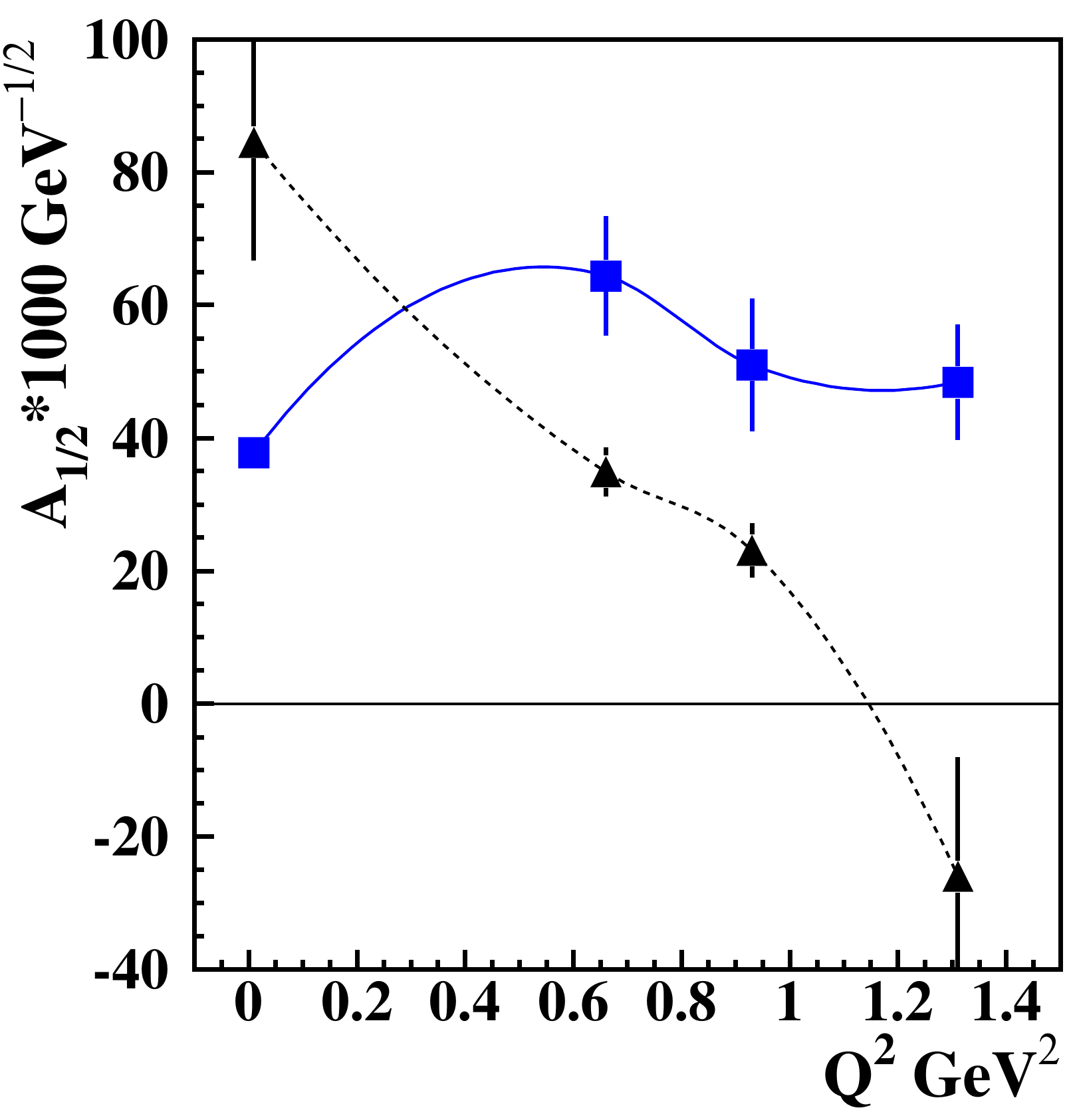}
\includegraphics[width=0.66\columnwidth]{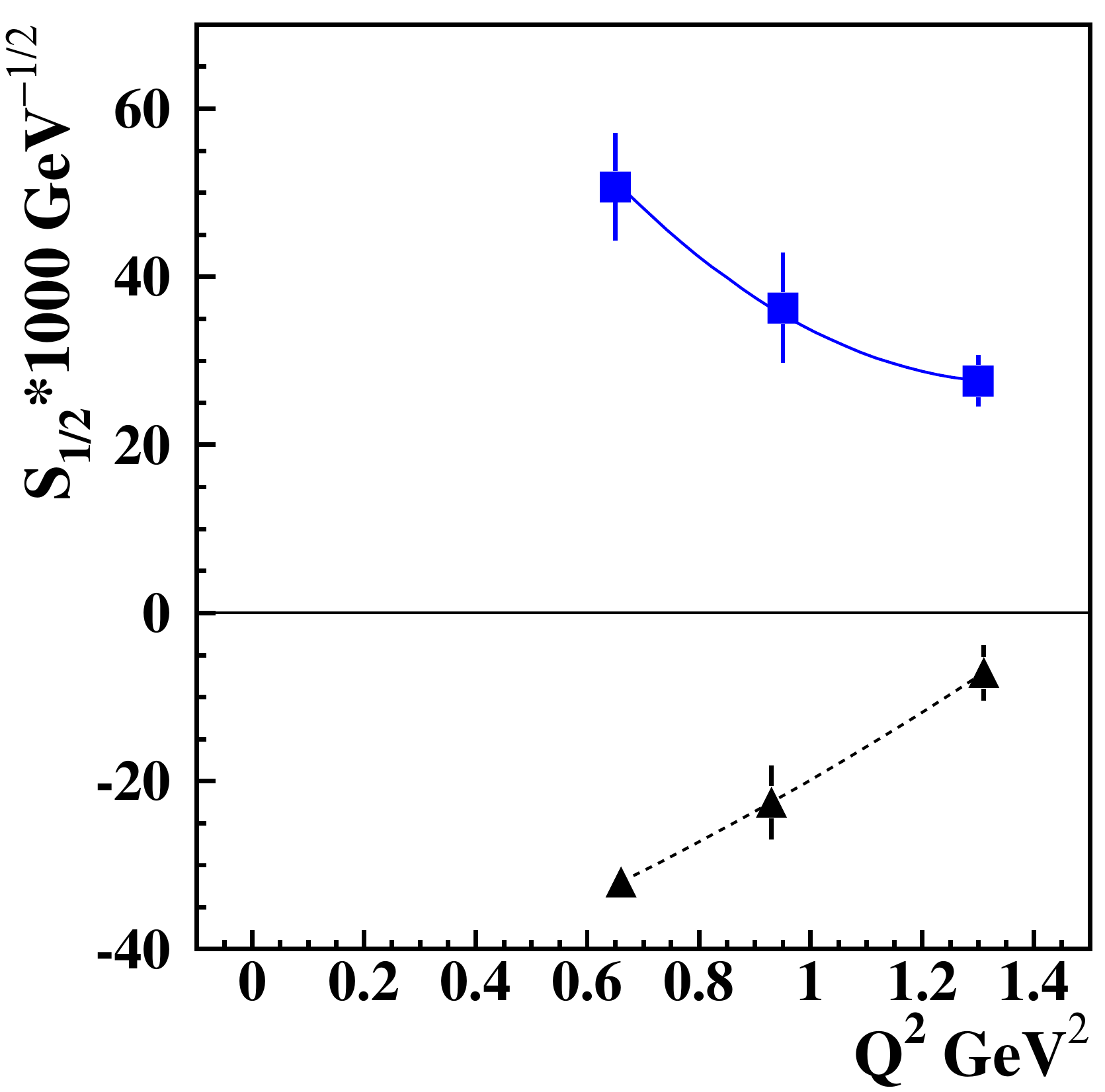}
\includegraphics[width=0.66\columnwidth]{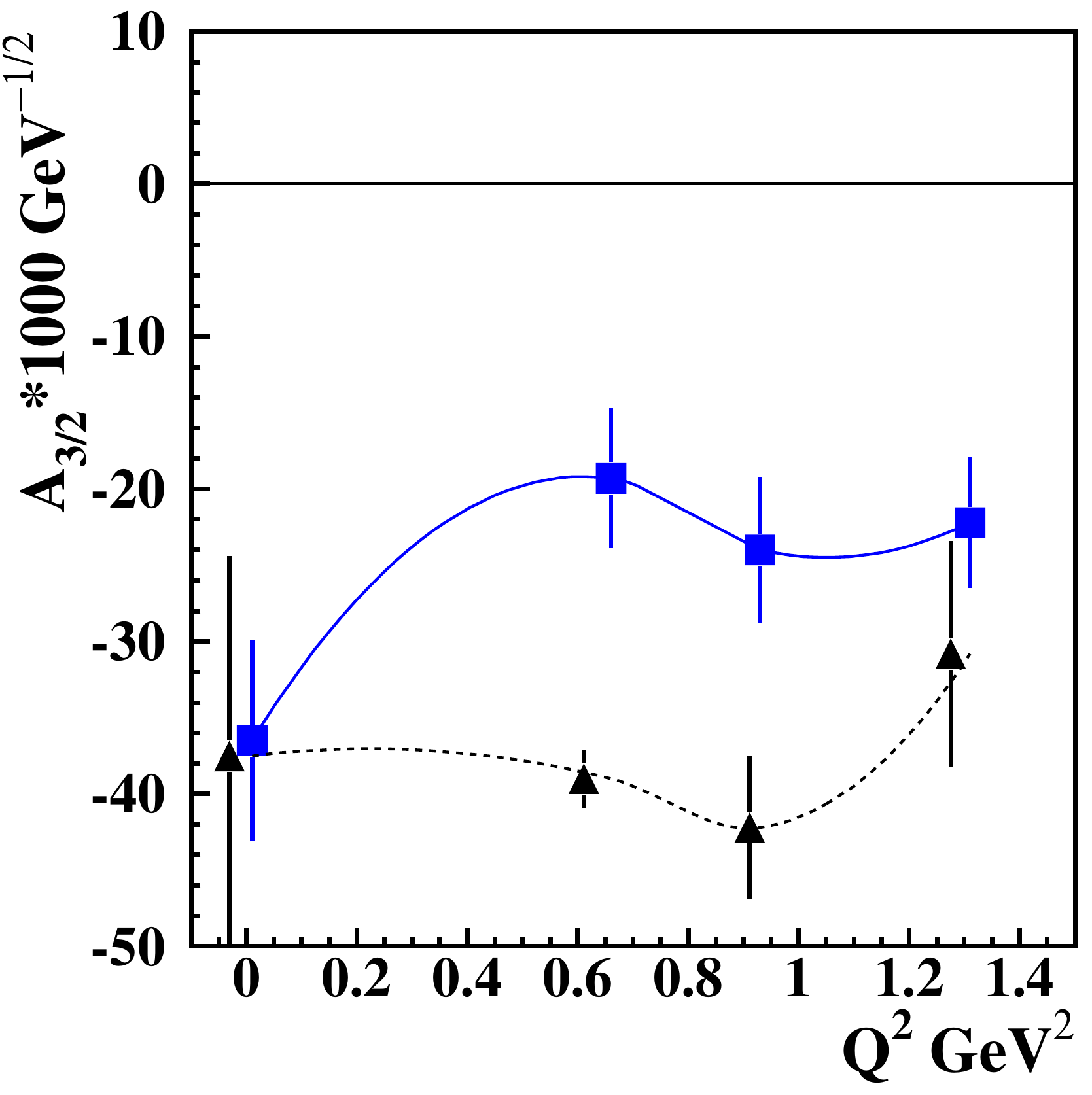}
\vspace{-0.3cm}
\caption{(Color Online) Comparison between the photo- and electrocouplings of the $N(1720)3/2^+$ (black triangles
  connected by dashed lines) and the new $N'(1720)3/2^+$ (blue squares connected by solid lines) obtained from the
  CLAS $\pi^+\pi^-p$ photo- and electroproduction data \cite{Ri03,Gol19}.}
\label{conv_miss}
\end{center}
\end{figure*}

\begin{table*}
\begin{center}
\caption{Masses and hadronic decay widths of the $N(1720)3/2^+$ and $N'(1720)3/2^+$ resonances to the $\pi \Delta$ and 
$\rho p$ final states determined as the overlap between the parameter ranges from independent fits of the $\pi^+\pi^-p$
photo- and electroproduction data~\cite{Ri03,Gol19}.}
\label{hadr_missconv}
\vspace{2mm}
\begin{tabular}{|c|c|c|c|c|} \hline
Resonance        & Mass,       & $N^*$ total width, & Branching fraction            & Branching fraction         \\
states              & GeV          & MeV                        &  for decays to $\pi\Delta$ &  for decays to $\rho p$ \\
\hline
$ N(1720)3/2^+$ & 1.743-1.753 &       114 $\pm$ 6       & 38-53\%                        &   31-46\%                     \\
\hline
$N'(1720)3/2^+$ & 1.715-1.735 &       120 $\pm$  6      & 47-62\%                         &    4-10\%                     \\
\hline
\end{tabular}
\end{center}
\end{table*}

Note that the decay widths of the well-known $N(1720)3/2^+$ resonance into the $\pi \Delta$ and $\rho p$ final
states depend considerably on the implementation of the new $N'(1720)3/2^+$ state. Accounting for only the
well-known resonances results in contradictory values for the $N(1720)3/2^+$ decays into the $\rho p$ final state
inferred either from the independent photo- or electroproduction data fits with more than a factor of four difference 
(see Table~\ref{hadrp13phel}). This makes it impossible to describe both the $\pi^+\pi^-p$ photo- and electroproduction
cross sections with $Q^2$-independent resonance masses, as well as total and partial hadronic decay widths, accounting 
for only the well-known resonances. 

After implementation of the new $N^{'}(1720)3/2^+$ resonance, a successful description of all nine 1-fold differential
$\gamma_{r,v}p \to \pi^+\pi^-p$  photo-/electroproduction cross sections has been achieved. The total, $\pi\Delta$,
and $\rho p$ hadronic decay widths of all resonances in the third resonance region as inferred from the fits at
different $Q^2$-bins remain $Q^2$-independent (see Table \ref{hadr_phot_miss}) over the entire range of $Q^2$
up to 1.5~GeV$^2$ that is covered by the measurements~\cite{Ri03,Gol19}. This supports the existence of the new
$N'(1720)3/2^+$ resonance. Indeed, if the implementation of this (or any) new baryon state was unphysical, then it 
would not be possible to reproduce the $\pi^+\pi^-p$ photo-/electroproduction data in a wide $Q^2$-range with 
$Q^2$-independent decay widths because of the evolution of the non-resonant contributions with $Q^2$ observed
in the $\pi^+\pi^-p$ electroproduction data~\cite{Mo16a}. 

The electrocouplings of the new $N'(1720)3/2^+$ resonance were determined in independent data fits for $Q^2$ in the
range from 0.5 -- 1.5~GeV$^2$ within the three overlapping $W$-intervals: 1.61 -- 1.71~GeV, 1.66 -- 1.76~GeV, and 
1.71 -- 1.81~GeV (see Fig.~\ref{new_elecph}). The non-resonant contributions in the three $W$-intervals are different, 
while the extracted $N'(1720)3/2^+$ electrocouplings agree within the uncertainties, which underlines the credible 
extraction of the electrocouplings. Furthermore, the $N'(1720)3/2^+$ mass, as well as the total and partial decay 
widths into the $\pi \Delta$ and $\rho p$ final states obtained from the fits in the three $W$-intervals, are also 
consistent, which further supports the existence of this new state.

Comparisons between the photo-/electroexcitation amplitudes of the $N(1720)3/2^+$ state and the new $N'(1720)3/2^+$ 
state determined from the CLAS $\pi^+\pi^-p$ photo-/electroproduction data~\cite{Ri03,Gol19} are shown in 
Fig.~\ref{conv_miss}. The transverse $A_{1/2}$ amplitude of the $N(1720)3/2^+$ resonance decreases with $Q^2$ more 
rapidly than for the new $N'(1720)3/2^+$ state. 

The contributions of the $N(1720)3/2^+$ and $N'(1720)3/2^+$ resonances to the fully integrated $\pi^+\pi^-p$ 
photo-/electroproduction cross sections are shown in Fig.~\ref{integ_newres}. As $Q^2$ increases the contributions 
from the $N'(1720)3/2^+$ become more pronounced relative to the $N(1720)3/2^+$. Both resonances are more visible in 
the electroproduction data compared to photoproduction. The sizable increase of the non-resonant contributions seen 
in the $\pi^+\pi^-p$ photoproduction data reduces the relative contributions from these resonances. 

Nevertheless, the combined studies of the $\pi^+\pi^-p$ photo-/electroproduction data are critical in order to validate 
the contributions from both the $N(1720)3/2^+$ and $N'(1720)3/2^+$ resonances. In the analyzed data set, it is only at 
the photon point that the contribution from the $N(1720)3/2^+$ is larger than that of the new $N'(1720)3/2^+$. The branching 
fraction range for the $N(1720)3/2^+$ decay from the photoproduction data into the $\rho p$ final state, $>$19\%, is imposed 
by the behavior of the high-mass part of the $\pi^+\pi^-$ mass distribution. This established range makes it impossible to 
simultaneously describe the $\pi^+p$ and $\pi^+\pi^-$ invariant mass distributions in the electroproduction data assuming 
only the contribution from the $N(1720)3/2^+$ resonance. This includes the pronounced $\Delta^{++}$ peaks seen in the 
$\pi^+p$ mass distributions and the absence of the $\rho$ peak in the $\pi^+\pi^-$ mass distributions (see 
Fig.~\ref{1diff_newres}). In order to reproduce the $\Delta^{++}$ peaks seen in the $\pi^+p$ mass distributions without including 
the $N'(1720)3/2^+$ state, the $N(1720)3/2^+$ decay widths to the $\rho p$ final state would have to be more than a factor of 
four smaller in electroproduction compared with the values established in photoproduction. When a new $N'(1720)3/2^+$ resonance 
is implemented, the $\Delta^{++}$ peaks in the electroproduction data in the $\pi^+p$ mass distributions can be well described 
by the contributions from the new $N'(1720)3/2^+$ state, which has only minor ($<$13\%) hadronic decays to the $\rho p$ final 
state. A rapid decrease of the $A_{1/2}$ electrocoupling of the $N(1720)3/2^+$ with $Q^2$ (see Fig.~\ref{conv_miss}) 
allows for the description of the $\pi^+\pi^-$ invariant mass distributions both in the photo- and electroproduction 
data, reproducing the high-mass part without the $\rho$ peak in the electroproduction reaction.

The masses, total decay widths, and branching fractions for the decays of these resonances into $\pi  \Delta$ and $\rho p$ final states
listed in Table~\ref{hadr_missconv} were evaluated as the overlap between the parameter ranges from independent fits of the 
$\pi^+\pi^-p$ photo- and electroproduction data. The new $N'(1720)3/2^+$  decays mostly into the $\pi\Delta$ final state, 
while the $N(1720)3/2^+$ decay widths into the $\pi\Delta$ and $\rho p$ final states are comparable. The contributions from 
the $N(1720)3/2^+$ and the new $N'(1720)3/2^+$ resonances are well separated in the $\pi^+\pi^-p$ photo-/electroproduction 
data analyses despite the close masses and the same spin-parity of these states. Different patterns for the decays 
into the $\pi\Delta$ and $\rho p$ final states and the different $Q^2$-evolution of the resonance electrocouplings allow 
us to disentangle their contributions. These differences can be seen in the combined studies of $\pi^+\pi^-p$ photo- 
and electroproduction, but they are elusive in the previous studies of the two-body meson-baryon channels, as well as 
the $\pi\pi N$ channels limited to photoproduction data only. Note that a global coupled-channel analysis of the exclusive 
meson photo- and hadroproduction data~\cite{lee13} has revealed evidence for two nucleon resonances of $J^P=3/2^+$ and 
$I$=1/2 for $W$ from 1.7 -- 1.8~GeV, supporting our observation of both the $N(1720)3/2^+$ and the new $N'(1720)3/2^+$ 
states.

\section{Shedding Light on the Nature of New Baryon States} 
\label{physimpact}

The discovery of several new resonances in the global multi-channel analysis of exclusive meson photoproduction data
\cite{Bu17} is consistent with the pattern from approximate SU(6) spin-flavor symmetry in the generation of the $N^*$
spectrum. Most of the states predicted in the $[70,2^+]$ multiplet have been observed. Two of them, the $N(1880)1/2^+$
and $N(1900)3/2^+$ with a 4-star status, and three others with a lower rating, are included in the PDG~\cite{Rpp18}.
However, one of the $[70,2^+]$ multiplet states of $J^P=3/2^+$ and isospin $I$=1/2~\cite{Klempt12} remains elusive.
Is it possible that the new $N'(1720)3/2^+$ resonance established in our analyses is this expected state? In order to
obtain an answer, we have estimated the mass of this state from the SU(6) symmetry pattern for the masses of the
nucleon resonances in the $[70,2^+]$ and $[70,1^-]$ multiplets. There are two resonances in the $[70,2^+]$ multiplet
with a total quark spin $S_q$=1/2: the still unknown state of $J^P=3/2^+$, isospin $I$=1/2 and the $N(1860)5/2^+$.
Four other states of $S_q$=3/2 are the $N(1880)1/2^+$, $N(1900)3/2^+$, $N(2000)5/2^+$, and $N(1900)7/2^+$
resonances. Their average mass is equal to 1.955~GeV. We assume that the difference between the average mass values
for the resonances of $S_q$=1/2 and $S_q$=3/2, $\Delta M(S_{3/2}-S_{1/2})$, in the $[70,2^+]$ multiplet is the same
as in the $[70,1^-]$. For the $[70,1^-]$ multiplet, $\Delta M(S_{3/2}-S_{1/2})=0.16$~GeV is obtained by averaging the
differences between the masses of the resonances $N(1650)1/2^-$, $N(1535)1/2^-$ and $N(1700)3/2^-$, $N(1520)3/2^-$ 
with $S_q$=1/2 and $S_q$=3/2. Hence, the average mass for the resonances of $S_q$=1/2, $M^{av}_{S=1/2}$, in the 
$[70,2^+]$ multiplet can be estimated as:
\begin{equation}
\begin{split}
M^{av}_{S=1/2}=M^{av}_{S=3/2}-\Delta M(S_{3/2}-S_{1/2}) \\
= 1.955\, {\rm GeV} - 0.16\, {\rm GeV} = 1.795\, {\rm GeV}.
\end{split} 
\end{equation}
The mass of the $N(1880)1/2^+$ state is smaller by $\Delta M$=0.075~GeV than $M^{av}_{S=3/2}$=1.955~GeV for the four
resonances of $S_q=3/2$ in the $[70,2^+]$ multiplet. Assuming the same $\Delta M$ for the $S_q$=1/2 doublet of
resonances in the $[70,2^+]$ multiplet, the mass of the lightest resonance of $S_q$=1/2, $M_{3/2^{+}}$, can be evaluated
as:
\begin{equation}
\begin{split}
M_{3/2^+ }=M^{av}_{S=1/2}-\Delta M \\
= 1.795\, {\rm GeV} - 0.075\, {\rm GeV} = 1.72\, {\rm GeV}.
\end{split}
\end{equation}
The estimated value of $M_{3/2^{+}}$ is in good agreement with the mass of the new $N'(1720)3/2^+$ resonance
from our analysis (see Table~\ref{hadr_missconv}), which makes plausible the assignment of the state as the lightest
resonance in the $[70,2^+]$ multiplet of $J^P=3/2^+$ and isospin $I$=1/2.

A variety of quark models predict two close resonances of $J^P=3/2^+$ and $I$=1/2, consistent with those seen in
our analysis. The interacting quark-diquark~\cite{San15} and the hypercentral constituent quark model
\cite{Giannini:2015zia} predict two states of $J^P=3/2^+$ and $I$=1/2 in the mass range from 1.7 -- 1.8~GeV. The
conceptually different chiral quark-soliton model~\cite{hyun} with parameters fit to the baryon masses in the octet and
decuplet predicts a resonance of $J^P=3/2^+$, $I$=1/2 in addition to the $N(1720)3/2^+$ as a member of the
27-SU(3)-baryon multiplet. The computed mass of this state of $1718.6 \pm 7.4$~MeV is consistent with the mass of
the new $N'(1720)3/2^+$ state (see Table~\ref{hadr_missconv}). The results on the $Q^2$-evolution of the
$N(1720)3/2^+$ and $N'(1720)3/2^+$ resonance electrocouplings have become available from our analysis for the first
time (see Fig.~\ref{conv_miss}). Confronting our findings with the quark model expectations will shed light on the missing
resonance nature, elucidating the peculiar features of strong QCD that have made these states elusive for such a long time.

\section{Summary}
\label{summ}

The analysis of the CLAS $\pi^+\pi^-p$ photo-/electroproduction data~\cite{Ri03,Gol19} has been carried out for $W$ 
from 1.6 -- 1.8~GeV and for $Q^2$ from 0 -- 1.5~GeV$^2$ with the goal to establish the nucleon resonances in the
third resonance region contributing to this channel. Accounting for only the well-established resonances results in 
more than a factor of four difference for the decay branching fractions of the $N(1720)3/2^+$ resonance into the 
$\rho p$ final state as inferred from independent fits of the $\pi^+\pi^-p$ photo-/electroproduction data (see
Table~\ref{hadrp13phel}). This contradiction makes it impossible to describe both the photo- and electroproduction
data unless the contributions from a still unobserved resonance are added.

After implementation of the $N'(1720)3/2^+$ resonance with photo-/electrocouplings, mass, and decay widths fit to
the CLAS data~\cite{Ri03,Gol19} (see Table~\ref{hadr_missconv} and Fig.~\ref{conv_miss}), a successful description of
the $\pi^+\pi^-p$ photo-/electroproduction data is achieved with $Q^2$-independent masses and total and partial decay
widths into the $\pi\Delta$ and $\rho p$ final states of all contributing resonances in the third resonance region.
Moreover, the photo-/electrocouplings and hadronic decay widths of all contributing resonances coincide within their
uncertainties determined from the independent fits in three overlapping $W$-intervals (see Fig.~\ref{new_elecph}). The 
evolution with $Q^2$ of the non-resonant contributions to $\pi^+\pi^-p$ electroproduction observed in the previous CLAS 
data analysis~\cite{Mo16a} makes it unlikely that the implementation of the $N'(1720)3/2^+$ resonance can serve 
as an effective way to describe the non-resonant contributions beyond the scope of the JM model and, therefore, these results 
support the existence of the new $N'(1720)3/2^+$ state. A manifestation of the new $N'(1720)3/2^+$ baryon state 
was also found in an independent global coupled-channel analysis of the exclusive meson photo- and hadroproduction data
\cite{lee13}, which also revealed evidence for two nearby resonances of $J^P=3/2^+$ and $I$=1/2 for $W$ from
1.7 -- 1.8~GeV.

The first results on the $Q^2$-evolution of the photo-/electroexcitation amplitudes of the missing baryon states 
have become available for the $N'(1720)3/2^+$. Confronting these results with the quark model predictions will shed 
light on the nature of the missing resonances. In the future, the observation of the new $N'(1720)3/2^+$ state will 
be also checked in the analysis of the recent high quality $\pi^+\pi^-p$ electroproduction data from CLAS
\cite{Is16,Fe18,Trivedi:2018rgo} in the $Q^2$ range from 0.4 -- 5.0~GeV$^2$. These data will provide the $TT$ and 
$LT$ interference structure functions~\cite{Trivedi:2018rgo} allowing for improvement in the evaluation of the 
resonance electrocouplings. 

Additional data on $\pi^+\pi^-p$ electroproduction at $Q^2 < 0.4$~GeV$^2$ are needed in order to explore the range of 
photon virtualities in Fig.~\ref{integ_newres} where the transition occurs from $N(1720)3/2^+$ dominance in photoproduction 
to $N'(1720)3/2^+$ dominance in electroproduction. The measurement of beam, target, and beam-target asymmetries will be very 
helpful for the extraction of the resonance electrocoupling, in particular at $Q^2 < 0.4$~GeV$^2$ where the non-resonant 
contributions become increasingly important as $Q^2$ goes to zero. The studies of $\pi\pi N$ photoproduction off protons 
with neutral hadrons in the final state at ELSA and MAMI will shed light on the manifestation of $N'(1720)3/2^+$ resonance in 
different exclusive $\pi\pi N$ channels needed for further confirmation of the existence of the new state. Our results on the 
mass, total decay width, and photo-/electrocouplings of the $N'(1720)3/2^+$ will guide the search for the manifestation of this state 
in other meson photo- and electroproduction channels, such as $KY$, $\omega p$, $\phi p$, $\pi\eta N$. The studies of exclusive $\pi\pi N$ 
hadroproduction planned at JPARC~\cite{Hicks} will allow for the exploration of the manifestation of the $N'(1720)3/2^+$ 
resonance in such reactions and to independently establish from hadroproduction data the $N(1720)3/2^+$ and $N'(1720)3/2^+$ 
decay widths to the $\pi\Delta$ and $\rho p$ final states.
 
\section{Acknowledgments}

We express our gratitude for valuable theoretical support by I.G. Aznauryan, T-S.H. Lee, C.D. Roberts, E. Santopinto, and
A.P. Szczepaniak. We would like to acknowledge the outstanding efforts of the staff of the Accelerator and the Physics
Divisions at Jefferson Lab that made the experiments possible. This work was supported in part by the U.S. Department of
Energy, the National Science Foundation, the University of Connecticut, Ohio University, the Skobeltsyn Institute of Nuclear
Physics, the Physics Department at Moscow State University, and the University of South Carolina. This material is based
upon work supported by the U.S. Department of Energy, Office of Science, Office of Nuclear Physics under contract
DE-AC05-06OR23177. The U.S. Government retains a non-exclusive, paid-up, irrevocable, world-wide license to publish or
reproduce this manuscript for U.S. Government purposes.


\begin{thebibliography}{99}

\bibitem{Bu16} V.D. Burkert, Few Body Syst. {\bf 57}, 873 (2016).

\bibitem{Cr13} V. Crede and W. Roberts, Rep. Prog. Phys. {\bf 76}, 076301 (2013).

\bibitem{dgirl19} D.G. Ireland, E. Pasyuk, and I. Strakovsky, Prog. Part. Nucl. Phys., {\bf 111} 103752 (2020).

\bibitem{Kl17} E. Klempt {\it et al.}, EPJ Web Conf. {\bf 134}, 02002 (2017). 

\bibitem{Be17} R. Beck {\it et al.}, EPJ Web Conf. {\bf 134}, 02001 (2017).

\bibitem{Capst} S. Capstick and W. Roberts, Prog. Part. Nucl. Phys. {\bf 45}, S241 (2000).

\bibitem{Giannini:2015zia} M.M. Giannini and E. Santopinto, Chin. J. Phys. {\bf 53}, 020301 (2015).

\bibitem{hyun} G.-S. Yang and H.-C. Kim, PTEP, 093D01 (2019).

\bibitem{Klempt12} E. Klempt and B. Ch. Metsch, Eur. Phys. J. A{\bf 48}, 127 (2012).

\bibitem{Du12} J.J. Dudek and R.G. Edwards, Phys. Rev. D {\bf 85}, 054016 (2012).

\bibitem{Ro11} H.L.L. Roberts {\it et al.}, Few Body Syst. {\bf 51}, 1 (2011).

\bibitem{Cr19s} Chen Chen {\it et al.}, Phys. Rev. D {\bf 100}, 054016 (2019).

\bibitem{Baz14} A. Bazavov {\it et al.}, Phys. Rev. Lett. {\bf 113}, 072001 (2014).

\bibitem{Capst1} S. Capstick, Nucl. Phys. Proc. Suppl. {\bf 45}, 241 (2000).

\bibitem{Bu17} A.V. Anisovich {\it et al.}, Phys. Rev. Lett. {\bf 119}, 062004 (2017).

\bibitem{Brad06} R.K. Bradford {\it et al.}  {\it (CLAS Collaboration)}, Phys. Rev. C {\bf 73}, 035202 (2006).

\bibitem{Brad07} R.K. Bradford {\it et al.} {\it (CLAS Collaboration)}, Phys. Rev. C {\bf 75}, 035205 (2007).

\bibitem{McC09} M.E. McCracken {\it et al.} {\it (CLAS Collaboration)}, Phys. Rev. C {\bf 81}, 025201 (2010).

\bibitem{Dey10} B. Dey {\it et al.} {\it (CLAS Collaboration)}, Phys. Rev. C {\bf 82}, 025202 (2010).

\bibitem{Rpp18} M. Tanabashi {\it et al.} {\it (Particle Data Group)}, Phys. Rev. D {\bf 98}, 03001 (2018).

\bibitem{Mo19} V.I. Mokeev, arXiv:1909.08746 [nucl-ex].

\bibitem{Mo16b} V.I. Mokeev {\it et al.}, EPJ Web Conf. {\bf 113}, 01013 (2016).

\bibitem{Ri03} M. Ripani {\it et al.} {\it (CLAS Collaboration)}, Phys. Rev. Lett. {\bf 91}, 022002 (2003).

\bibitem{Bu12} I.G. Aznauryan and V. D. Burkert, Prog. Nucl. Part. Phys. {\bf 67}, 1 (2012).

\bibitem{Is16} E.L. Isupov {\it et al.} {\it (CLAS Collaboration)}, Phys. Rev. C {\bf 96}, 025209 (2017).

\bibitem{Gol19} E.N. Golovatch {\it et al.} {\it (CLAS Collaboration)}, Phys. Lett. B {\bf 788}, 371 (2019).

\bibitem{lee13} H. Kamano {\it et al.}, Phys. Rev. {\bf C 88}, 035209 (2013).

\bibitem{San15} E. Santopinto and J. Ferretti, Phys. Rev. {\bf C 92}, 025202 (2015).

\bibitem{Mokeev:2008iw} V.I. Mokeev {\it et al.}, Phys. Rev. {\bf C 80}, 045212  (2009).

\bibitem{Mokeev:2012vsa} V.I. Mokeev {\it et al.} {\it (CLAS Collaboration)}, Phys. Rev. {\bf C 86}, 035203 (2012).

\bibitem{Mo16a} V.I. Mokeev {\it et al.}, Phys. Rev. C {\bf 93}, 054016 (2016).

\bibitem{Man92} D.M. Manley and E.M. Salesky, Phys. Rev. D {\bf 45}, 4002 (1992). 

\bibitem{Ast} A.N. Hiller Blin {\it et al.}, Phys. Rev. {\bf C 100}, 035201 (2019).

\bibitem{webIsupov} Nucleon Resonance Photo-/Electrocouplings Determined from Analyses of
Experimental Data on Exclusive Meson Electroproduction off Protons,\\
\url{https://userweb.jlab.org/~mokeev/resonance\_electrocouplings/}

\bibitem{Fe18} G.V. Fedotov {\it et al.}  {\it (CLAS Collaboration)}, Phys. Rev. {\bf C 98}, 025203 (2018).

\bibitem{Trivedi:2018rgo} A. Trivedi, Few Body Syst. {\bf 60}, 45 (2019).

\bibitem{Hicks} K.H. Hicks and H. Sako, J-PARC experiment E45, “3-Body Hadronic Reactions for New Aspects of
  Baryon Spectroscopy”, 2012, http://j-parc.jp/researcher/Hadron/en/Proposal\_e.html

\end{thebibliography}
\end{document}